\documentclass[sn-mathphys-num]{sn-jnl}

\usepackage[style=numeric-comp, sorting=none]{biblatex}
\usepackage{graphicx}%
\usepackage{multirow}%
\usepackage{amsmath,amssymb,amsfonts}%
\usepackage{amsthm}%
\usepackage{mathrsfs}%
\usepackage[title]{appendix}%
\usepackage{xcolor}%
\usepackage{textcomp}%
\usepackage{manyfoot}%
\usepackage{booktabs}%
\usepackage{algorithm}%
\usepackage{algorithmicx}%
\usepackage{algpseudocode}%
\usepackage{listings}
\usepackage{caption}
\usepackage{subcaption}
\usepackage{quantikz}
\usepackage{enumitem}
\newtheorem{theorem}{Theorem}

\usepackage{array}

\newcolumntype{C}[1]{>{\centering\arraybackslash}p{#1}}

\begin{document}

\title[Deterministic Quantum Search for Arbitrary Initial Success Probabilities]{Deterministic Quantum Search for Arbitrary Initial Success Probabilities}



\author{\fnm{Harishankar} \sur{Mishra}}\email{harishankarm@cdac.in}

\author{\fnm{Asvija} \sur{Balasubramanyam}}\email{asvijab@cdac.in}

\author*[ ]{\fnm{Gudapati Naresh} \sur{Raghava}}\email{nareshraghava@cdac.in}

\affil{\centering{\orgdiv{Quantum Technology Group}}\\ \orgname{Centre for Development of Advanced Computing (C-DAC)}\\ \orgaddress{\city{Bangalore}, \country{India}}}


\abstract{Unstructured search remains one of the significant challenges in computer science, as classical search algorithms become increasingly impractical for large-scale systems due to their linear time complexity. Quantum algorithms, notably Grover's algorithm, leverage quantum parallelism to achieve quadratic speedup over classical approaches. Its generalization, the amplitude amplification algorithm, extends this advantage to a broader class of search problems. However, both algorithms are inherently probabilistic and fail to guarantee success with certainty, also under the standard implementation of these algorithms, they offer no quantum advantage when the initial success probability exceeds 0.5. This work presents a deterministic quantum search algorithm that operates effectively for arbitrary initial success probabilities, providing guaranteed success in searching target states. The proposed approach introduces at most one additional iteration beyond the optimal count required by probabilistic quantum search algorithms. The algorithm preserves the quadratic speedup characteristic of quantum search methods. Additionally, an approach is proposed for the exact search of multiple target states within a bounded number of steps. A complete circuit-level implementation of the proposed algorithm is also presented.}
\maketitle
\section{Introduction}\label{sec1}
    Quantum computing \cite{QC} is an advanced field of computer science that leverages the unique principles of quantum mechanics to solve problems beyond the capabilities of even the most powerful classical computers. Quantum computing is a computational framework that utilizes the principles of quantum mechanics, such as superposition, interference and entanglement, to solve certain problems more efficiently as compared to classical computers. It utilizes qubits, which enable the evaluation of multiple inputs simultaneously, providing a significant advantage over classical computers in processing large scale data and complex computations.\\\\
    The unstructured database search is one of the challenging problems in computing, for which adopting quantum algorithmic approaches have been shown to provide significant advantages. This search problem involves exploring every element in the database to find the desired target. Classical deterministic search techniques generally solve this problem simply by iterating over each element one by one, to search for the target element, with the query complexity of  O$(N)$ \cite{cormen2009introduction}, where $N$ is the number of elements in the database. The time required to search increases linearly with respect to the size of the database, creating a bottleneck in many real world applications.\\ \\
    There are quantum algorithms developed to search in unstructured databases that offer speedups as compared to any existing classical approaches. These quantum search algorithms have shown potential in different fields like cryptography \cite{crypt, xu2024quantum, denisenko2019application, fernandez2024implementing, mandal2024implementing}, optimization \cite{opt, qopt, chakrabarty2017dynamic}, cryptanalysis \cite{cryptanalysis}, database searching\cite{db} and even in searching for solutions to computationally hard problems in areas such as artificial intelligence and machine learning \cite{aiml, innan2024variational}. The most notable among the quantum search algorithms, is the Grover’s algorithm \cite{GRO, grover1997quantum}, introduced by Lov Grover in 1996. This algorithm offers a quadratic speedup for unstructured search problem compared to its classical counterparts. Grover’s algorithm can perform unstructured search in $O(\sqrt{N})$ query in a database consisting of $N$ elements, significantly reducing the number of operations required.
    \\\\
    While Grover's algorithm offers significant advantages in problems where searching through large number of data or exploring a vast solution space is required, it assumes that initially all the states are equally likely and the total number of states to be power of two. This limits the application of Grover's algorithm in the fields where there is an arbitrary number of states available or if the initial probability distribution of states is not uniform. To overcome these shortcomings, a more generalized form of Grover's algorithm known as quantum amplitude amplification algorithm was developed by Brassard et al. \cite{aa} back in 2002. This quantum amplitude amplification algorithm can be used to search for a target element in any arbitrary database of size $N$, while providing the same quadratic speedup as provided by the Grover's algorithm, maintaining the overall query complexity of O$\big(\sqrt{N}\big)$.\\\\
    However, despite their capabilities, Grover's and amplitude amplification algorithms are not without limitations. One of the challenges with these algorithms is their probabilistic nature. While these algorithms reduce the number of iterations needed to find the target, their outcome is probabilistic. Hence, these algorithms must be executed multiple times to increase the probability of finding the target state. Furthermore, when target states constitute more than $50\%$ of the total states in the database, Grover's algorithm and amplitude amplification, reduce the probability of finding the target states compared to the initial success probability, thereby offering no quantum advantage.\\\\
    For many real-world applications, a deterministic outcome is desirable, particularly when the cost of failure is high. This is especially important in domains where complete precision is necessary, like security-related cryptography \cite{aest}, optimization \cite{opt} and search problems. There have been multiple approaches to mitigate the probabilistic nature of these algorithms. These methods primarily achieve deterministic success by modifying the phase rotations within the Grover operator, either through precise phase matching or multi-phase sequences (explanation and working of Grover operators will be discussed in detail in Section. \ref{sec2}) \cite{long2001grover, apqaa, dqs, dgsro, li2007phase, yoder2014fixed, toyama2008multiphase} or having a near deterministic approach without phase design in oracle or diffusion operators \cite{ndro}.\\\\
    In contrast, the approach proposed in this work follows a fundamentally different strategy. Instead of modifying the internal phase structure of the Grover operator, the effective initial success probability is adjusted through an ancilla-assisted construction. In standard amplitude amplification, the state evolution is confined to a two-dimensional subspace spanned by the marked and unmarked components, with a rotation angle determined by the initial success probability $p$. Deterministic success requires this rotation reach $\pi/2$, which, in general, cannot be achieved for arbitrary $p$ using an integer number of iterations. Existing exact search methods address this limitation by modifying the Grover operator through phase engineering.\\\\
    The proposed approach achieves the same objective by modifying the effective projection onto the two-dimensional subspace, rather than altering the rotation operator itself. By embedding the system into an enlarged Hilbert space using an ancilla qubit, an effective success probability $p'$ is realized such that the corresponding rotation angle leads to exact success after an integer number of Grover iterations. Importantly, this construction preserves the standard black-box nature of the oracle and avoids the need for precise phase calibration within the Grover operator. This provides an alternative realization of deterministic amplitude amplification that shifts the design complexity from operator phase engineering to state preparation.\\\\
    This article seeks to address the inherent probabilistic nature of Grover's and amplitude amplification algorithms by developing a deterministic quantum search approach. It proposes a method to modify these algorithms to achieve deterministic outcomes while also addressing their limitation in the regime where the initial success probability exceeds $0.5$. Importantly, the proposed approach preserves the same order of query complexity as Grover's algorithm corresponding to the upper bound on the optimal number of oracle queries requiring at most one additional iteration compared to the standard Grover procedure.\\ \\
    This article will first explore the Grover's and amplitude amplification algorithm briefly, also analyzing their limitations (Section. \ref{sec2}). Later, the article discusses about an approach to eliminate the probabilistic behavior of these algorithms (Section. \ref{sec3}), mathematically derive the expression for this deterministic algorithm (Section. \ref{sec4}) and build the corresponding quantum circuit (Section. \ref{sec6}) for implementing this algorithm and discuss its complexity. Lastly, an additional use case is discussed for searching multiple target states in definite number of steps (Section. \ref{sec7}).


\section{Grover’s and Amplitude Amplification Algorithms : Principles and Limitations}\label{sec2}
    \subsection*{Principles and Working of Grover's Algorithm}
\begin{figure}
            \centering
                \begin{subfigure}[t]{0.45\textwidth}
                    \includegraphics[width=5cm]{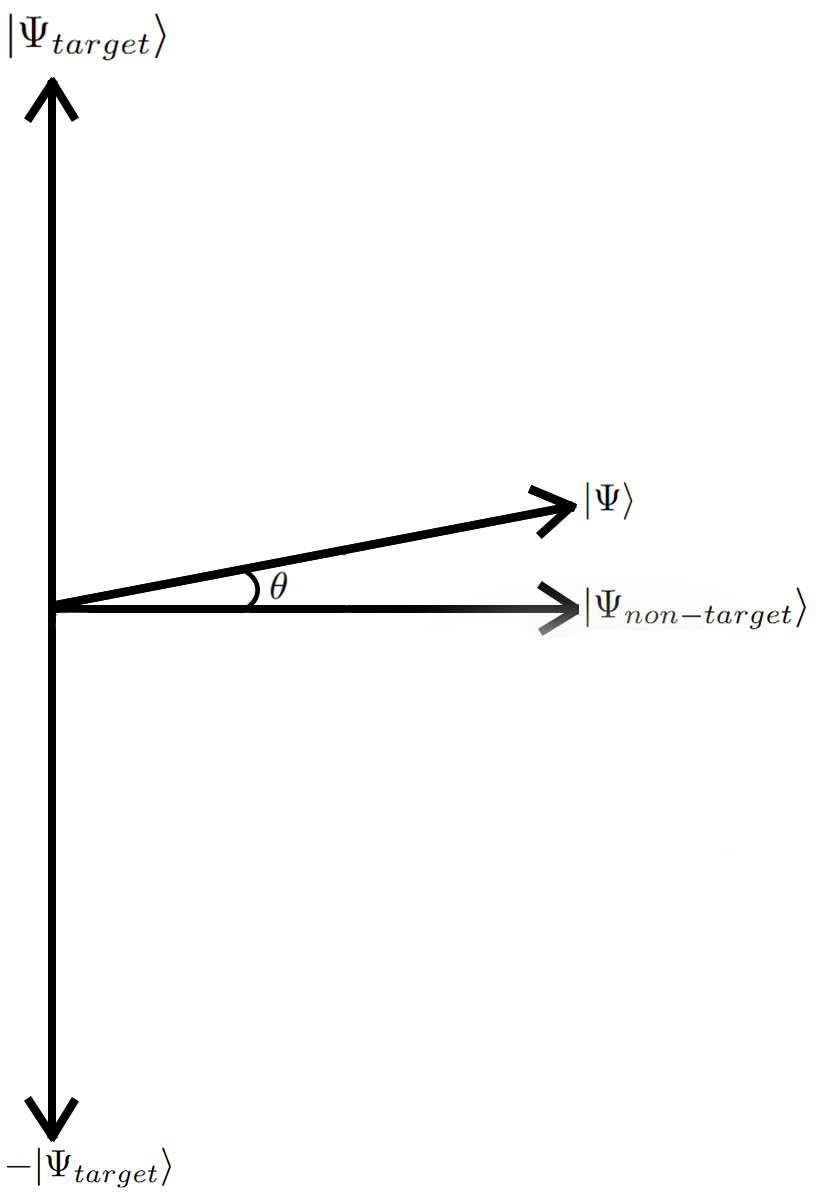}
                    \caption{Initial state of system, $|\Psi\rangle$,\\ where $\theta = sin^{-1}\Big(\sqrt{\frac{M}{N}}\Big)$.}
                    \label{fig:Initial_State}
                \end{subfigure}
                \begin{subfigure}[t]{0.45\textwidth}
                    \includegraphics[width=5cm]{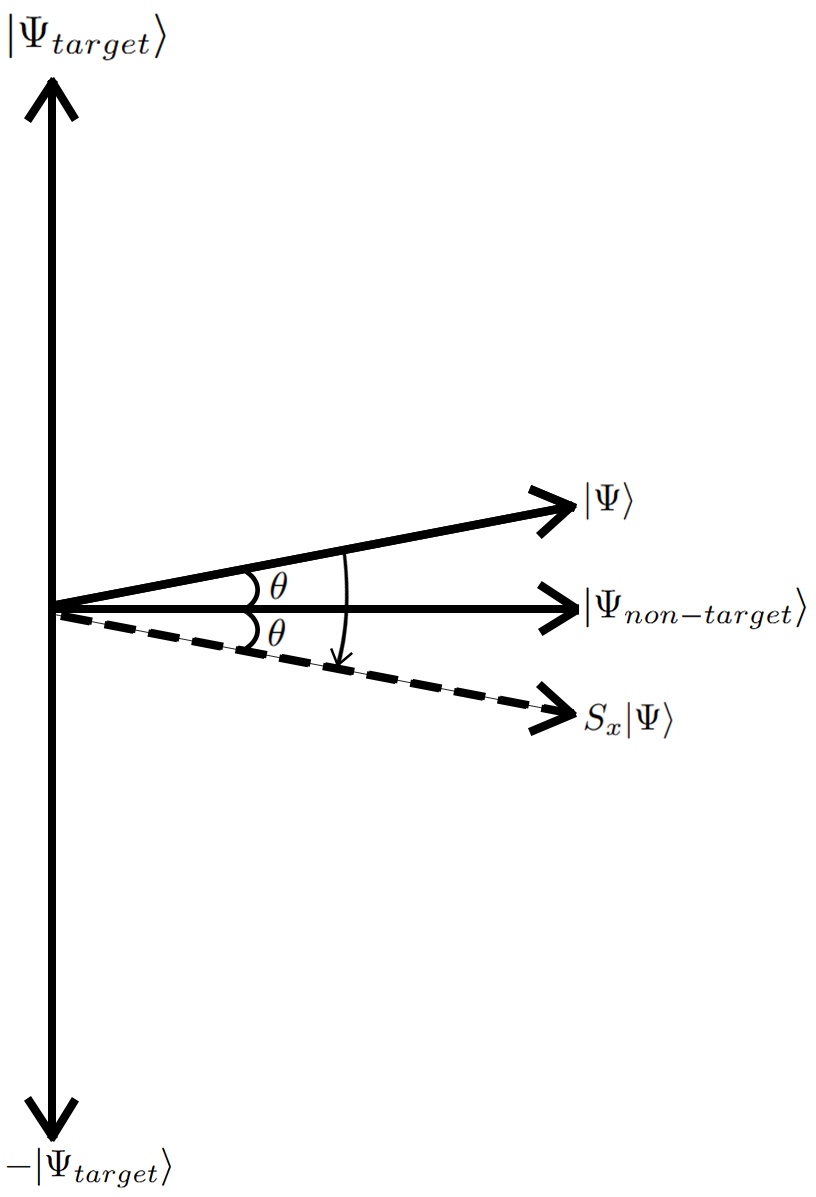}
                    \caption{Representation of the state after applying the oracle operator.}
                    \label{fig:Oracle_State}
                \end{subfigure}
            \vfill
                \begin{subfigure}[t]{0.45\textwidth}
                    \includegraphics[width=5cm]{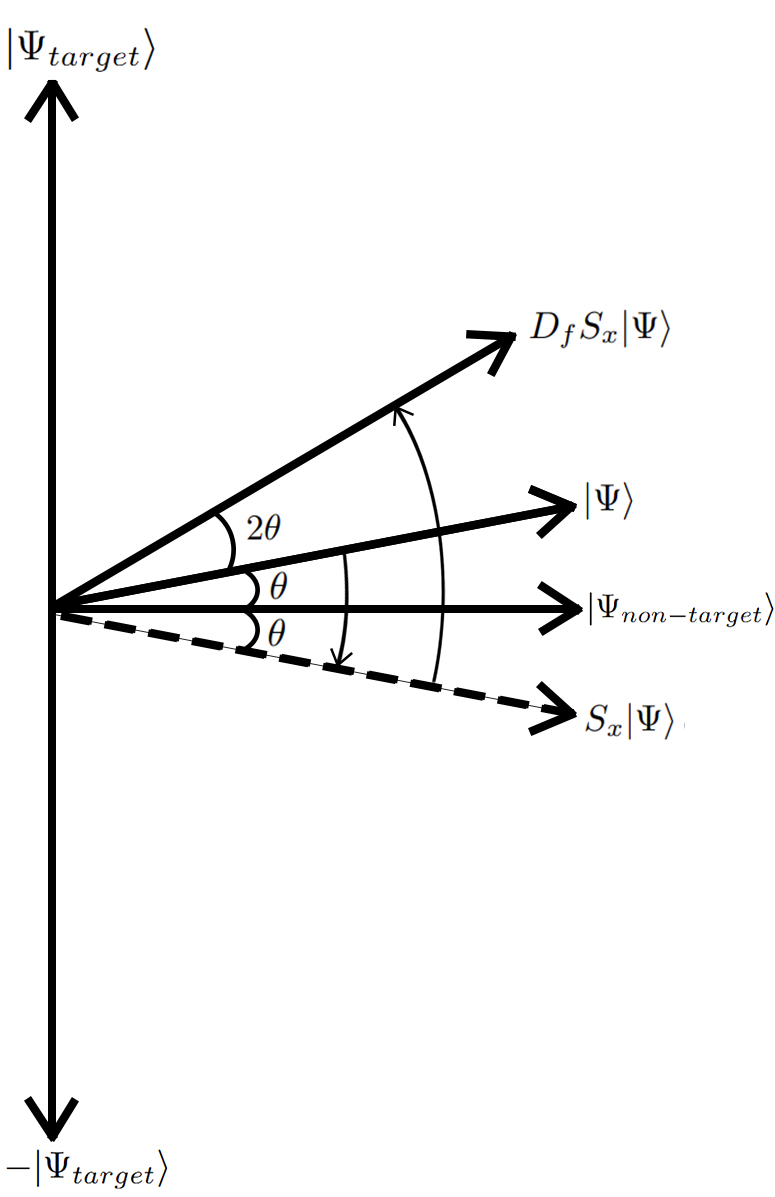}
                    \caption{Representation of the state after applying the diffusion operator.}
                    \label{fig:Diffusion_State}
                \end{subfigure}
        	    \begin{subfigure}[t]{0.45\textwidth}
                    {\includegraphics[width=5cm]{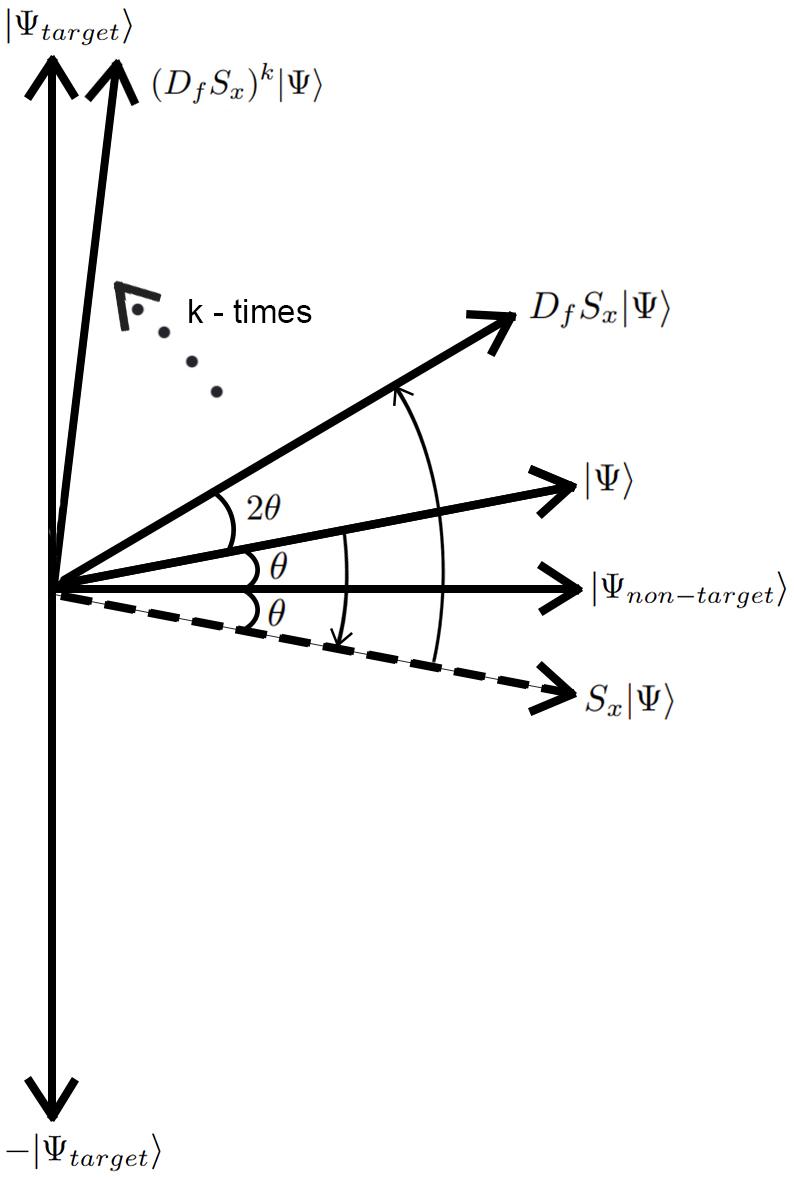}}
                    \caption{Representation of the state after applying the Grover's operator $k$ times}
                    \label{fig:Final_State}
        	    \end{subfigure}
        	\caption{Evolution of the state in Grover's algorithm.}
        	\label{fig:subfigures4}
        \end{figure}

        Grover’s algorithm was designed to find target states in an unsorted database. Alternatively, the problem statement can be understood as follows: consider a boolean function $f(x)$ that takes an input $x$ from $N$ possibilities and returns a boolean value of $0$ or $1$. The objective of Grover's algorithm is to find the input(s) $x$ (target states) for which the function's output is $1$ \cite{formal_des}. Classically, finding a specific item in an unordered database of $N$ entries requires checking each entry one by one, leading to overall query complexity of $O(N)$. Grover’s algorithm, however, uses quantum superposition and interference to search the database in significantly fewer steps than classical methods. The basic idea behind Grover's algorithm is as follows:
        \vspace{0.3cm}
        \begin{enumerate}
            \item The algorithm starts by creating a uniform quantum superposition of all possible states (i.e. all database entries) using a quantum register.
            \item It then performs a sequence of quantum operations, specifically the oracle and diffusion operator (collectively referred to as the Grover's operator), to increase the probability of measuring the target states.
            These operations are repeated  O$(\sqrt{N})$ times to maximize the probability of measuring the target states.
            \item Finally, a quantum measurement collapses the superposition into one of the possible states, with the target states being most likely.
        \end{enumerate}
        \vspace{0.3cm}
        The key quantum operations of the Grover’s operator are:
        \vspace{0.3cm}
        \begin{itemize}
            \item Oracle: A quantum operation that marks the target state(s), by adding a phase of $\pi$ to their probability amplitude.
            \item Diffusion operator: This operator amplifies the probability amplitudes of the marked target states while simultaneously decreasing the amplitudes of all other states, thereby increasing the likelihood of measuring a target state.
        \end{itemize}
        \vspace{0.3cm}
        The Grover's operator, $G$, is defined as:
        \vspace{0.3cm}
        \begin{equation}
            \label{eq:grover operator}
            G(|\Psi\rangle \otimes |y\rangle) = (H^{\otimes n}S_0H^{\otimes n} \otimes I)S_x\big(|\Psi\rangle \otimes |y\rangle\big)
        \end{equation}
        where:
        \vspace{0.3cm}
        \begin{itemize}
            \item $n$ is the number of qubits used to represent all $N = 2^n$ possible computational basis states.
            \item $|\Psi\rangle$ is an arbitrary quantum state.
            \item $H$ represents Hadamard operation.
            \item $S_x$ represents oracle operator, which applies a phase flip to the target states: $S_x(|\Psi\rangle \otimes |-\rangle) = -(|\Psi\rangle \otimes |-\rangle)$, if $|\Psi\rangle$ is a target state and $S_x(|\Psi\rangle \otimes |-\rangle) = (|\Psi\rangle \otimes |-\rangle)$ otherwise (the detailed working of this oracle operator can be seen in Appendix. \ref{secA1}).
            \item $S_0$ applies the inversion about the $|0\rangle^{\otimes n}$ state, defined as $S_0|0\rangle^{\otimes n} = -|0\rangle^{\otimes n}$ and $S_0|\Psi\rangle = |\Psi\rangle$ for all other computational basis state $|\Psi\rangle \neq |0\rangle^{\otimes n}$.
        \end{itemize}
        \vspace{0.3cm}
        Together, the $H^{\otimes n}S_0H^{\otimes n}$ operator is called the diffusion operator and let it be represented as $D_f$.
        The application of the Grover's algorithm on an arbitrary state can be understood as follows:\\\\
        Consider that there is a state $|\Psi\rangle$ as a uniform superposition of $N$ states, such that $N = 2^n$ for some arbitrary value of $n$. The state $|\Psi\rangle$ can be represented as:
        $$|\Psi\rangle = H^{\otimes n}|0\rangle$$
        $$|\Psi\rangle = \frac{1}{\sqrt{N}}\sum_{x = 0}^{N-1}|x\rangle$$
        This state can be represented as the superposition of two orthogonal states: $|\Psi_{target}\rangle$ (the uniform superposition of all $M$ target states) and $|\Psi_{non-target}\rangle$ (the uniform superposition of all $N-M$ non-target states). Hence, the state of the system can also be represented as:
        $$|\Psi\rangle = \sqrt{\frac{M}{N}}|\Psi_{target}\rangle + \sqrt{\frac{N - M}{N}}|\Psi_{non-target}\rangle$$
        where, $M$ is the number of target states. This state can also be written in a geometric representation:
        $$|\Psi\rangle = sin{\theta}|\Psi_{target}\rangle + cos{\theta}|\Psi_{non-target}\rangle$$
        where, $\theta$ is angle between the state $|\Psi\rangle$ and $|\Psi_{non-target}\rangle$ as shown in Fig. \ref{fig:Initial_State}. The initial success probability, $p$ (probability of obtaining a target state when $|\Psi\rangle$ is measured), is given by: 
        \begin{equation}
            \label{eq:p}
            p = \frac{M}{N} = sin^2(\theta)
        \end{equation}
        The objective is to perform operations on $|\Psi\rangle$ to maximize its amplitude along the target state $|\Psi_{target}\rangle$, thereby increasing the probability of measuring the desired output. This is achieved by successively applying the oracle ($S_x$) and the diffusion operator ($D_f$). The oracle applies a reflection about the non-target states $|\Psi_{non-target}\rangle$. This operation adds a phase of $\pi$ only to the component of $|\Psi\rangle$ along $|\Psi_{target}\rangle$ state, and act as an identity operator on the component along $|\Psi_{non-target}\rangle$. The operation of the oracle is illustrated in Fig. $\ref{fig:Oracle_State}$.\\\\
        Following the application of the oracle, the diffusion operator is applied to amplify the amplitude of the target states. This operator reflects the current state of the system about the initial uniform superposition state $|\Psi\rangle$ of the system. This operation can be represented as in Fig. $\ref{fig:Diffusion_State}$. The probability amplitude of the component of state $D_fS_x|\Psi\rangle$ along $|\Psi_{target}\rangle$ increases compared to that of the initial state $|\Psi\rangle$. To maximize the probability amplitude of the component of the state $|\Psi\rangle$ along $|\Psi_{target}\rangle$ to be (ideally) 1, the operator $D_fS_x$ (the Grover operator) is applied successively $k$ times, as depicted in Fig. \ref{fig:Final_State}.\\
        \\
        The physical realization of this mathematical evolution can be represented by a generic quantum circuit. Fig. \ref{fig:generalized Grover's} illustrates the standard schematic block diagram of Grover's algorithm, where an $n$-qubit data register is first prepared in a uniform superposition using Hadamard gates, followed by $k$ successive applications of the Grover operator ($D_f S_x$) prior to measurement.\\
        \begin{figure}[h]
            \centering
                \definecolor{myblue}{RGB}{173, 216, 230}
                \definecolor{myred}{RGB}{255, 182, 193}
                \definecolor{mygreen}{RGB}{144, 238, 144}
                \definecolor{mygrey}{RGB}{151, 151, 151}
                \fbox{\begin{quantikz}
                    \lstick{$\ket{0}^{\otimes n}$} 
                        & \qwbundle{n}
                        & \gate[style={fill = myred}]{H^{\otimes n}}
                        & \gate[style={fill=mygreen},label style=black, 2][3cm]{S_x}\gateinput{$x$}\gateoutput{$x$}
                        \gategroup[wires=2, steps=2, style={dashed, rounded corners, inner sep=6pt}, label style={label position=below, anchor=north, yshift=-0.3cm}]{$k$ iterations} 
                        & \gate[style={fill=mygreen},label style=black]{D_f}
                        & \meter[style={fill=mygrey},label style=black]{}\\
                    \lstick{$ancilla$ $\ket{-}$} 
                        &                       
                        &
                        & \gateinput{$y$}\gateoutput{$y\oplus f(x)$}
                        &
                        &
                \end{quantikz}}
            \caption{Generic quantum circuit schematic of the standard Grover's algorithm. The $n$-qubit quantum register is prepared in a uniform superposition via Hadamard gates ($H^{\otimes n}$), while an $ancilla$ qubit is initialized in the $|-\rangle$ state to facilitate phase kickback. The Grover operator, comprising the oracle ($S_x$) and the diffusion operator ($D_f$), is applied iteratively $k$ times to amplify the target states before measuring the data register.}
            \label{fig:generalized Grover's}
        \end{figure}\\
        After each iteration, the state of the system is rotated by an angle of $2\theta$ in the two-dimensional subspace spanned by $|\Psi_{target}\rangle$ and $|\Psi_{non-target}\rangle$. Thus, after $k$ iterations, the angle between the final state and the $|\Psi_{non-target}\rangle$ becomes $\theta' = k\times2\theta + \theta = (2k + 1)\theta$ and the probability of measuring a target state is then given by:
        \begin{equation}
            \label{eq:p_success}
            p_{success} = sin^2((2k + 1)\theta)
        \end{equation}\\
        The objective is to apply the Grover's operator multiple times to achieve $p_{success} = 1$. This implies that the angle between the final state $(D_fS_x)^k|\Psi\rangle$ and the $|\Psi_{non-target}\rangle$ state should be $\pi/2$. The number of iterations $k$ required to achieve this is given by
        \begin{equation}
            \label{eq:k}
            \implies k = \frac{\pi}{4sin^{-1}\Big(\sqrt{p}\Big)} - \frac{1}{2}
        \end{equation}
        If $\sqrt{M/N} << 1$ then $sin^{-1}(\sqrt{M/N}) \approx \sqrt{M/N}$, which allows for the following approximation for $k$:
        \begin{equation}
            \label{eq:quadratic speed up}
            k \approx \frac{\pi}{4}\sqrt{\frac{N}{M}} - \frac{1}{2}
        \end{equation}
        Eqn \ref{eq:quadratic speed up} shows that the overall query complexity is O($\sqrt{\frac{N}{M}}$), indicating a quadratic speedup compared to classical search algorithms. In standard Grover's algorithm, if the calculated value of $k$ turns out to be a decimal, it is rounded off to the nearest natural number.
    \subsection*{Amplitude Amplification Algorithm}
         Grover's algorithm works under the assumption that the initial state is uniform superposition of all possible states. This limitation can be addressed by using the amplitude amplification algorithm, which generalizes the principles of Grover’s algorithm. The overall operator of the amplitude amplification, $Q$, is given by :
        $$Q = -AS_0A^{-1}S_x$$
        where $S_x$ and $S_0$ are same as in Grover's algorithm as in Eqn. \ref{eq:grover operator} and $A$ is the state preparation operator such that $A|0\rangle = |\Psi\rangle$, where $|\Psi\rangle$ is any arbitrary initial state. Thus, amplitude amplification algorithm is a more generalized form of Grover's algorithm, with Grover's algorithm being a specialized case where the initial arbitrary state is uniform superposition of all possible states.
    \subsection*{Limitations}
\begin{figure}[H]
            \centering
            \fbox{\includegraphics[width=6cm]{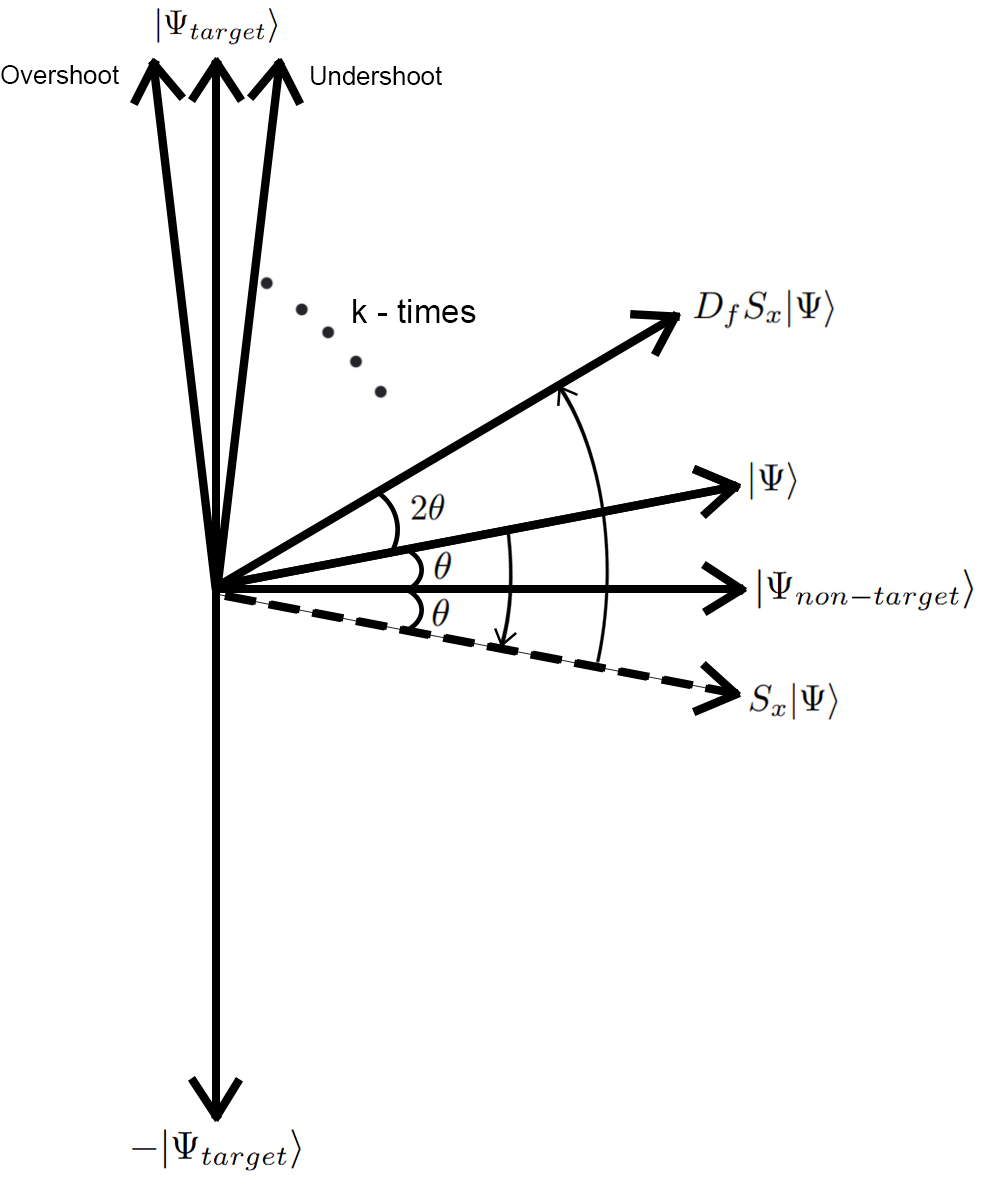}}
            \caption{State representation for undershoot or overshoot based on value chosen for number of iterations, $k$.}
            \label{fig:Error_State}
        \end{figure}

\begin{figure}[H]
            \centering
            \fbox{\includegraphics[width=7cm]{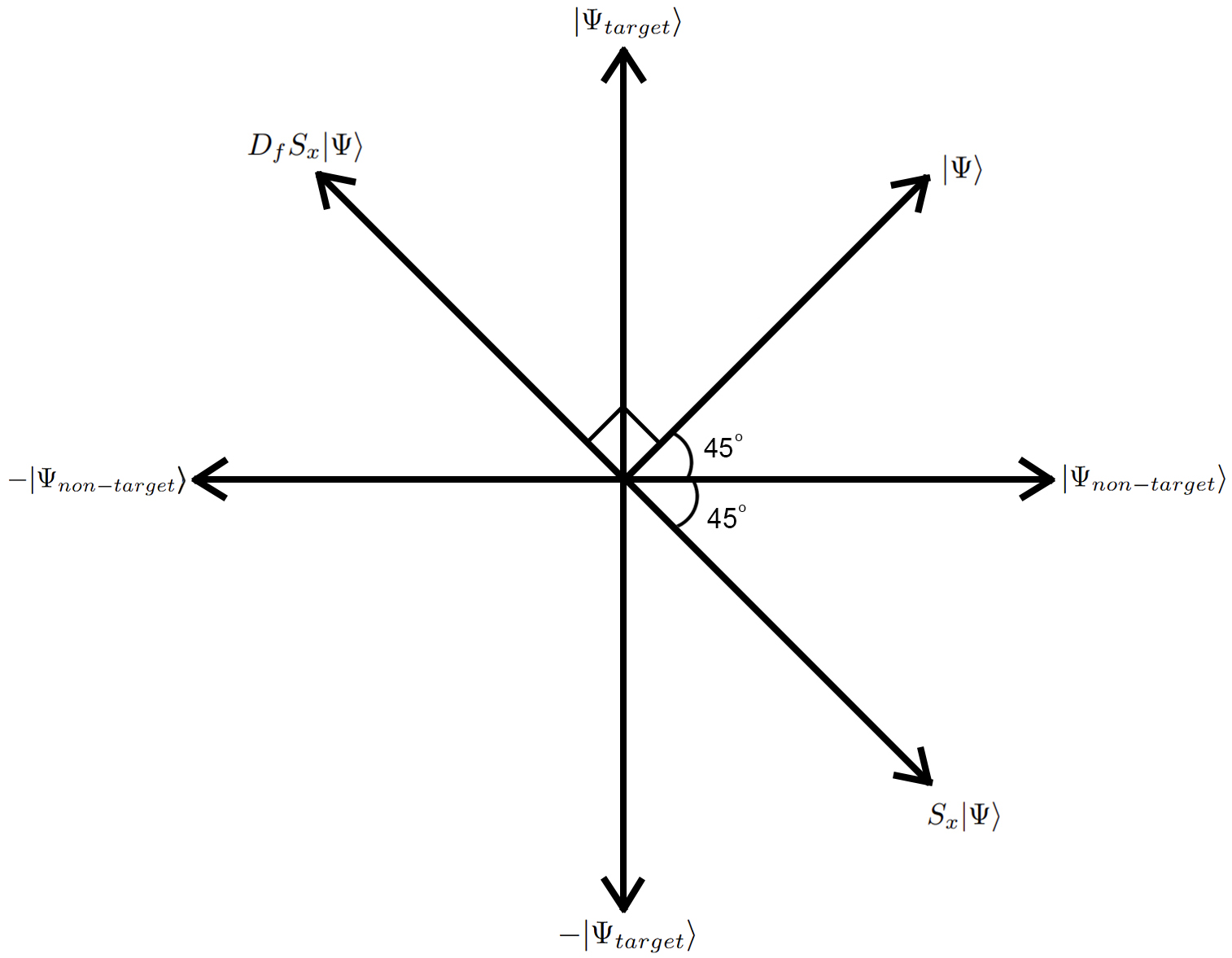}}
            \caption{Representation of states for initial success probability equals to $0.5$.}
            \label{fig:50}
        \end{figure}

        Despite its quadratic advantage, one notable limitation of these quantum search algorithms is that they are probabilistic. This probabilistic nature stems from non-fractional number of application of Grover's operator, which from Eqn. \ref{eq:k} is clearly not constrained to be a natural number. Thus, the computed value of $k$ is rounded and either the lower or upper bound is selected accordingly. However, if the lower bound ($k_{l} < k$) is chosen, then the final angle between the state vector and the non-target subspace is given by
        $$2(k_{l} + 1)\theta \ < \ 2(k + 1)\theta = \frac{\pi}{2}$$
        which results in undershoot and the final state of the system does not fully overlap with $|\Psi_{target}\rangle$ as shown in Fig. \ref{fig:Error_State}. Similarly if the upper bound of $k$ is chosen ($k_u > k$), the corresponding angle satisfies
        $$2(k_{u} + 1)\theta \  > \ 2(k + 1)\theta = \frac{\pi}{2}$$
        leading to an overshoot, causing the state to rotate beyond the $|\Psi_{target}\rangle$. In both the cases, the resulting success probability is less than $1$, as the state vector of the system in these cases has non-zero vector component along $|\Psi_{non-target}\rangle$. The quantum measurement collapses the superposition of states, introduces an inherent randomness into the result. While the Grover's and amplitude amplification algorithms increases the probability of measuring the correct solution, these algorithm does not guarantee that the correct answer will be found on the first execution of the algorithm. In practice, this implies that these algorithms must be repeated multiple times to enhance the probability of measuring the correct solution.\\
        \noindent Another limitation inherent to Grover's and amplitude amplification algorithms is that when the target state makes up for more than half of the initial state $|\Psi\rangle$, these algorithms instead of increasing the probability of measuring the target states, decreases its probability. This could be understood as, if the initial success probability $p$ is greater than $0.5$, the number of iterations calculated from Eqn. \ref{eq:k} is less than $0.5$. This value is typically rounded to $0$, which implies that there is no quantum advantage since the Grover's operator would not be applied. Where as if the upper bound of $k$ is considered, i.e. $1$ then the probability of success after the application of oracle and diffusion operator turns out to be $0.5$ (Eqn. \ref{eq:p_success}), which is same as the initial state. This is represented as in Fig. $\ref{fig:50}$.\\
        \noindent Now, if the initial probability of success is greater than $0.5$ then the number of iterations calculated from Eqn. \ref{eq:k} is less than $0.5$, which is rounded to $0$, hence there is no quantum advantage as the number of times the operator is needed to be applied is $0$. Even if the operator is applied for $1$ time the final success probability will always be less than the initial success probability, if the initial success probability is more than $0.5$. This can be proven as follows:\\\\
        Given the initial success probability, $p = sin^2\theta$ (from Eqn. \ref{eq:p}), the case where $p \geq 1/2 $ is considered. This implies $sin^2\theta \geq \frac{1}{2}$.\\\\
        The purpose is to show that $p \geq p_{success}$  after one iteration, i.e., $sin
        ^2\theta \geq sin^2(3\theta)$.\\
        Starting with:
        \begin{align*}
            &sin^2\theta \ge \frac{1}{2}\\
            &2sin^2\theta - 1 \ge 0
        \end{align*}
        So it can be written as,
        \begin{align*}
            0 \ge (2sin^2\theta - 1)(sin^2\theta - 1) 
        \end{align*}
            as $sin^2(\theta) \le 1$ for any value of $\theta$,
        \begin{align*}
            0 \ge 2sin^4\theta - 3sin^2\theta + 1
        \end{align*}
        multiplying the inequality by 8,
        \begin{align*}
            0 \ge 16sin^4\theta - 24sin^2\theta + 8
        \end{align*}
        adding $1$ on both sides,
        \begin{align*}
            1 \ge 16sin^4\theta - 24sin^2\theta + 9
        \end{align*}
        multiplying both sides by $sin^2\theta$ ($sin^2\theta \neq 0$ as $sin^2\theta \ge \frac{1}{2}$):
        \begin{align*}
            sin^2\theta \ge 16&sin^6\theta - 24sin^4\theta + 9sin^2\theta \\
            sin^2\theta& \ge (3sin\theta - 4sin^3\theta)^2 \\
            &sin^2\theta \ge sin^2(3\theta) \\
        \end{align*}
        This demonstrates that the initial success probability, given by $sin^2\theta$ (Eqn. \ref{eq:p}), is always greater than or equal to the success probability after a single iteration, given by $sin^2(3\theta)$ (Eqn. \ref{eq:p_success}). In such a scenario, Grover's and amplitude amplification algorithms offer no quantum advantage over classical methods.\\\\
        The observation that, for $p \geq 0.5$, standard amplitude amplification does not yield advantage under the integer-iteration prescription highlights a limitation of directly applying Grover's algorithm or amplitude amplification in this regime. This behavior arises because the optimal number of iterations is less than one, preventing effective amplitude amplification. The proposed deterministic approach addresses this by modifying the effective initial success probability, thereby enabling a non-zero integer number of iterations that achieves deterministic success. Consequently, the method extends the applicability of amplitude amplification to regimes where the standard formulation is not effective.
\section{Towards Deterministic Quantum Search}\label{sec3}
    This section focuses on the foundation of the deterministic quantum search algorithm, which mitigates the probabilistic nature of quantum search techniques and provides an advantage even when the initial success probability of measuring the target state(s) is greater than $0.5$. The proposed approach uses Grover’s and amplitude amplification algorithms to achieve deterministic quantum search. To begin with, the working of this algorithm can be understood by analyzing the variation of theoretical success probability with respect to the initial success probability ($p$) of these standard quantum search algorithms, as given by $p_{success}$ in Eqn. \ref{eq:p_success}.
    \begin{figure}[H]
        \centering
        \fbox{\includegraphics[width=0.98\textwidth]{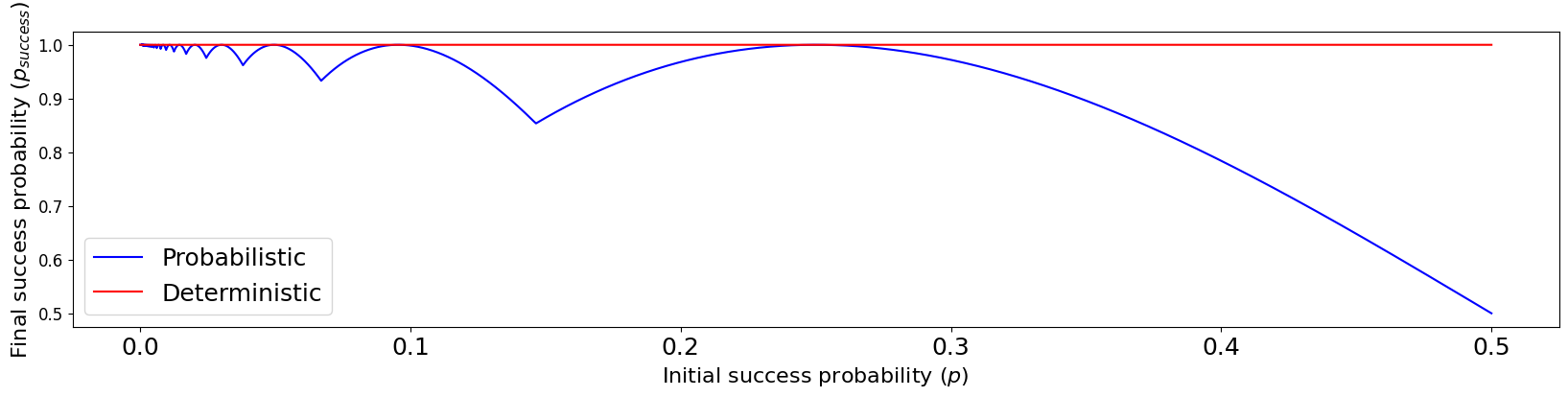}}
        \caption{Success probability comparison between probabilistic methods (Grover’s and amplitude amplification algorithms) and deterministic outcome\\(success probability = 1).}
        \label{fig:theory_success_1}
    \end{figure}
    \noindent Fig. \ref{fig:theory_success_1} illustrates the success probability of Grover’s and amplitude amplification algorithms as a function of the initial success probability ($p$) and also highlights the deterministic outcome. The graph is shown only for $p \le 0.5$, because, for $p > 0.5$, the number of iterations ($k$) for these algorithms becomes less than $0.5$, which rounds to nearest integer $0$. As a result, Grover’s and amplitude amplification algorithms does not offer quantum advantage  in this regime, as explained in Section \ref{sec2}. However, the deterministic approach proposed in this section, yields deterministic results even when $p > 0.5$. The working of this deterministic search algorithm can be understood by scaling Fig. \ref{fig:theory_success_1} as depicted in Fig. \ref{fig:theory_success_2}. 
    \begin{figure}[H]
        \centering
        \fbox{\includegraphics[width=0.98\textwidth]{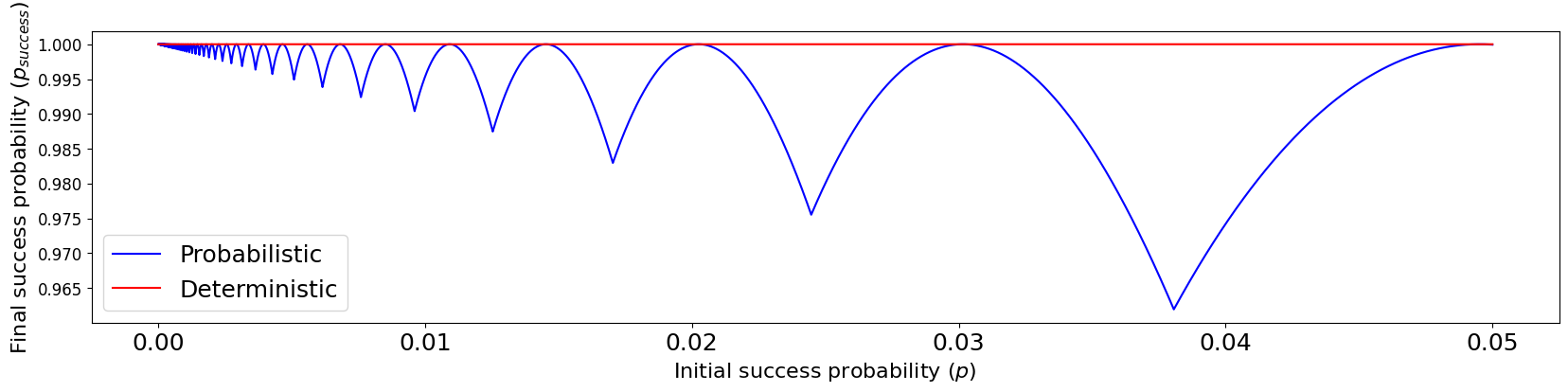}}
        \caption {Success probability comparison between probabilistic methods (Grover’s and amplitude amplification algorithms) and deterministic outcome\\scaled version of Fig. \ref{fig:theory_success_1}.}
        \label{fig:theory_success_2}
    \end{figure}
    \noindent From Fig. \ref{fig:theory_success_2}, it is evident that when the target state is present in the search space (i.e., $p > 0$), for any value of initial success probability $p$, there always exist lower initial success probabilities ($p' < p$) for which $p_{success}$ is $1$ (the detailed mathematical explanation is provided in Appendix \ref{secA3}). As explained in Section \ref{sec2}, $p_{success}$ is $1$ only when the number of iterations ($k$) is a natural number. This characteristic is vital for the development of a deterministic quantum search algorithm proposed in this section.\\\\
    The deterministic approach defines an effective target state ($|\Psi_{new\_target}\rangle$) as the tensor product of the $ancilla$ state $|0\rangle$ and the original target state ($|\Psi_{target}\rangle$), i.e.,
    $$|\Psi_{new\_target}\rangle = |0\rangle \otimes |\Psi_{target}\rangle$$
    This construction modifies the effective initial success probability from $p$ to $p'$. Among all admissible values of $p'$, the one closest to $p$ that yields an integer iteration count is chosen for optimality. This is achieved by preparing the ancilla qubit in a suitable superposition state such that the probability of measuring $|\Psi_{new\_target}\rangle$ equals $p'$. A deterministic result is obtained by applying the oracle and diffusion operator to search $|\Psi_{new\_target}\rangle$. The oracle marks the $|\Psi_{target}\rangle$ states only when the $ancilla$ qubit is in state $|0\rangle$.  This behavior is achieved by invoking the standard black-box oracle within a controlled configuration using the $ancilla$ qubit, where the control state corresponds to $|0\rangle$.
    \vspace{0.3cm}\\
    When the initial success probability of measuring the target state is $0.5$ or more, in such cases the $ancilla$ qubit is initialized in such a way that the probability of measuring $|\Psi_{new\_target}\rangle$ becomes $0.25$ (the value of $p'$ corresponding to $k' = \lceil k \rceil = 1$ in this regime of $p \geq 0.5$). In this case with just one iteration the superposition of all target states is obtained. Measurement of all qubits except for the $ancilla$ qubit collapses the state of the system into one of the target states. The detailed explanation on how to achieve this deterministic search is mentioned in Section \ref{sec4}.
\section{Algorithmic Framework}\label{sec4}
    The key idea is that when the number of iterations ($k$) is not a natural number in Grover's or amplitude amplification algorithms, the algorithm initializes the $ancilla$ qubit in a specific superposition instead of the conventional $|-\rangle$ state. Assuming that the initial success probability $p$ is known or can be estimated. This specific initialization ensures that the recalculated number of iterations ($k'$), based on the probability of measuring the target states in conjunction with the $ancilla$ qubit being in the state $|0\rangle$, becomes the closest natural number greater than $k$.\\
    Consider an arbitrary initial state $|\Psi\rangle$, which is represented as $$|\Psi\rangle = \sum_{i=0}^{S-1}S_i|s_i\rangle$$
    where $S$ is the total number of states, $\{s_0, \ s_1,\ s_2,...,\ s_{S-1}\}$ are the states and $S_i$ is the probability amplitude of state $|s_i\rangle$
    This state can be represented in terms of target states ($|\Psi_{target}\rangle$) and non target states ($|\Psi_{non\_target}\rangle $) as:\\
    $$|\Psi\rangle = \sqrt{p} |\Psi_{target}\rangle + \sqrt{1-p} |\Psi_{non\_target}\rangle $$\\
    Here $p$ denotes the probability of measuring a $|\Psi_{target}\rangle$ when the initial state is measured.\\\\
    After initializing the $ancilla$ qubit, the state of the system changes to 
    \begin{equation}
        \label{eq: Psi_dash}
        |\Psi^{'}\rangle = \Big(\alpha|0\rangle + \beta|1\rangle \Big) \otimes \Big(\sqrt{p} |\Psi_{target}\rangle + \sqrt{1-p} |\Psi_{non\_target}\rangle\Big)
    \end{equation}
    where the value of $\alpha$ and $\beta$ depends on how the $ancilla$ qubit is initialized, with $|\alpha|^2 + |\beta|^2 = 1$. The new target state for the deterministic approach is defined as :
    \begin{equation}
        \label{eq: target_new}
        |\Psi_{new\_target}\rangle = |0\rangle \otimes |\Psi_{target}\rangle
    \end{equation}
    This results in an initial success probability of measuring this new target state $|\Psi_{new\_target}\rangle$ as:
    \begin{equation}
        \label{eq: p_dash}
        p' = p|\alpha|^2
    \end{equation}
    In the deterministic approach, the number of iterations is desired to be a natural number. For optimal query complexity when $k$ is not a natural number, $k'$ is considered as :
    \begin{equation} 
        \label{eq: k_and_kexact}
        k' = [k] + 1
    \end{equation}
    where $[A]$ denotes the integral part of $A$.\\\\
    The $ancilla$ qubit is initialized in such a way that the number of iterations calculated based on Eqn. \ref{eq:k} given by,
    \begin{equation}
        \label{eq: kpX}
        k' = \frac{\pi}{4sin^{-1}\Big(\sqrt{p|\alpha|^2}\Big)} - \frac{1}{2}
    \end{equation}
    is a natural number. From Eqn. \ref{eq: p_dash} and Eqn. \ref{eq: kpX}, the new initial probability of success can be written as
    \begin{equation}
        \label{eq: k_dash_p_dash}
        p' = sin^2\Big(\frac{\pi}{4(k' + 1/2)}\Big)
    \end{equation}
    Since Eqn. \ref{eq: k_and_kexact} dictates that $k' > k$, and the number of iterations is inversely related to the square root of the success probability (Eqn. \ref{eq: k_dash_p_dash}), this implies that the new effective initial success probability $p|\alpha|^2$ is less than the original initial success probability $p$ ($p|\alpha|^2 < p$). This requires $|\alpha|^2 < 1$, which is always possible to satisfy by choosing an appropriate value for the probability amplitude $\alpha$ during the initialization of the $ancilla$ qubit in its superposition state.\\\\
    To obtain the desired value of $\alpha$, the $ancilla$ qubit is initialized using the $R_y$ quantum operator. The matrix representation of this gate is given by
    $$R_y(\phi) = 
    \begin{bmatrix}
        cos(\frac{\phi}{2}) & -sin(\frac{\phi}{2})\\ \\
        sin(\frac{\phi}{2}) & cos(\frac{\phi}{2})
    \end{bmatrix}$$
    Applying this gate to the $|0\rangle$ state results in the superposition $cos(\frac{\phi}{2}) |0\rangle + sin(\frac{\phi}{2}) |1\rangle$. Comparing this with the general state of the $ancilla$ qubit (Eqn. \ref{eq: Psi_dash}), we have $$cos^2\Big(\frac{\phi}{2}\Big) = |\alpha|^2$$
    From Eqn. \ref{eq: p_dash}, ($p' = p|\alpha|^2$), we can substitute $|\alpha|^2$:
    $$cos^2\Big(\frac{\phi}{2}\Big)p = p'$$
    Solving for $\phi$, we get:
    \begin{equation}
        \label{eq: theta}
        \phi = 2 cos^{-1}\Big(\sqrt{\frac{p'}{p}}\Big)
    \end{equation}
    \vspace{0.3cm}\\
    Thus, the $ancilla$ qubit is initialized with $R_y(\phi)$ gate using the angle from Eqn. \ref{eq: theta}.\\
    The overall flow of this deterministic search algorithm is as follows:
    \vspace{0.3cm}
    \begin{description}[
        font=\normalfont,   
        labelwidth=4em,     
        leftmargin=4em,   
        labelsep=0em      
    ]
        \item[Step 01:] Determine the number of iterations ($k$) using the given initial success probability ($p$), as per Eqn. \ref{eq:k}.
        \item[Step 02:] If $k$ is a natural number, set $k' = k$ and proceed directly to Step 8.
        \item[Step 03:] (If $k$ is not a natural number) : Calculate the number of iterations, $k'$, based on the value of $k$, using Eqn. \ref{eq: k_and_kexact}.
        \item[Step 04:] Compute the new required initial success probability ($p'$) using Eqn. \ref{eq: k_dash_p_dash}.
        \item[Step 05:] Find the angle of rotation ($\phi$) for the $R_y$ operator from Eqn. \ref{eq: theta}.
        \item[Step 06:] Initialize the $ancilla$ qubit in superposition with $R_y(\phi)$ gate using the angle $\phi$.
        \item[Step 07:] Apply a controlled invocation of the oracle conditioned on the ancilla qubit.
        \item[Step 08:] Apply the diffusion operator to increase the amplitude of the marked states.
        \item[Step 09:] Repeat Step 07 and Step 08, $k'$ times.
        \item[Step 10:] Measure the final state to obtain one of the target states with certainty.
    \end{description}
    \vspace{1em}
    In the algorithmic sequence outlined above, the construction of the controlled oracle required for Step 07 is achieved by applying external gates to the $ancilla$ qubit while strictly treating the standard oracle as an unobservable black box. By applying specific single qubit operations, Pauli-$X$ and Hadamard gates to the $ancilla$ qubit immediately before and after the standard oracle query, the required phase kickback is conditionally triggered exclusively when the $ancilla$ is in the $|0\rangle$ state. This approach shows that the required controlled oracle operation can be achieved without modifying or accessing the internal structure of the oracle. The mathematical proof demonstrating the exact working of this controlled oracle achieved through external configuration is detailed in Appendix \ref{secA2}.

    \vspace{1em}
    \noindent A generic circuit layout of the deterministic quantum search algorithm is presented in Fig. \ref{fig:deterministic_circuit}. The circuit comprises three principal components: (i) initialization of the $ancilla$ qubit using an $R_y(\phi)$ rotation (ii) an oracle that marks the new target state $|0\rangle \otimes |\Psi_{\text{target}}\rangle$ and (iii) a diffusion operator. The oracle and the diffusion operators are applied $k'$ times to achieve deterministic amplification of the target state.
    \begin{figure}[H]
    \centering
        \definecolor{myblue}{RGB}{173, 216, 230}
        \definecolor{myred}{RGB}{255, 182, 193}
        \definecolor{mygreen}{RGB}{144, 238, 144}
        \definecolor{mygrey}{RGB}{151, 151, 151}
        \fbox{\begin{quantikz}
            \lstick{$\ket{0}^{\otimes n}$} 
                & \qwbundle{n}          
                & \gate[style={fill=mygreen},label style=black]{A}              
                &
                \gategroup[wires=2, steps=6, style={dashed, rounded corners, inner sep=6pt}, label style={label position=below, anchor=north, yshift=-0.3cm}]{$k'$ iterations} 
                &
                & \gate[style={fill=mygreen},label style=black, 2][3cm]{S_x}\gateinput{$x$}\gateoutput{$x$}            
                & 
                &
                & \gate[style={fill=mygreen},label style=black, 2]{D_f}           
                & \meter[style={fill=mygrey},label style=black]{} \\           
            \lstick{$ancilla$ $\ket{0}$} 
                &                       
                & \gate[style={fill=myblue},label style=black]{R_y(\phi)}      
                & \gate[style={fill=myred, circle, inner sep=-2pt},label style=black]{+}
                & \gate[style={fill=myred},label style=black]{H}
                & \gateinput{$y$}\gateoutput{$y\oplus f(x)$}
                & \gate[style={fill=myred},label style=black]{H}
                & \gate[style={fill=myred, circle, inner sep=-2pt},label style=black]{+}
                &                       
                &               
        \end{quantikz}}
    \caption{Generalized quantum circuit for the Deterministic Quantum Search algorithm. The $ancilla$ qubit is initialized using an $R_y(\phi)$ rotation to obtain the required initial success probability, while the $n$ qubit register is initialized using operator $A$. The oracle $S_x$  is invoked conditionally using the ancilla qubit, introducing phase marking of the target states when the ancilla is in the state $|0\rangle$. The diffusion operator $D_f$ acts on the joint state of both the $ancilla$ and $n$ qubit register. The core search subroutine (oracle $S_x$ and diffusion operator $D_f$) is applied $k'$ times to ensure a deterministic measurement outcome.}
    \label{fig:deterministic_circuit}
\end{figure}
    The algorithmic representation of the deterministic quantum search is shown in Algorithm \ref{algo1}:\\
    \begin{algorithm}[H]
        \caption{Deterministic Quantum Search}\label{algo1}
        \begin{algorithmic}[1]
        \Require $0 < p < 1$
        \State $k \Leftarrow \frac{\pi}{4sin^{-1}(\sqrt{p})} - \frac{1}{2}$
        \If{$k \% 1 == 0$}\label{algln2}
                \State $k' \Leftarrow k$
        \Else
                \State $k' \Leftarrow floor(k) + 1$
                \State $p' \Leftarrow sin^2\Big(\frac{\pi}{4(k' + 1/2)}\Big)$
                \State $\phi \Leftarrow 2 cos^{-1}\Big(\sqrt{\frac{p'}{p}}\Big)$
                \State Initialize $ancilla$ qubit with $R_y(\phi)$
                \State Construct the controlled oracle for marking new target states.
        \EndIf
        \While{$k' \neq 0$}
                \State Apply Oracle
                \State Apply Diffusion operator
                \State $k' \Leftarrow k' - 1$
        \EndWhile
        \State Measure the final state
        \end{algorithmic}
    \end{algorithm}
    \noindent Once the number of iterations ($k'$) is determined and the system is prepared in the corresponding initial state (through $ancilla$ qubit initialization), a deterministic search result is obtained after $k'$ iterations. This yields a deterministic outcome within at most one additional iteration (as per Eqn. \ref{eq: k_and_kexact}) compared to standard Grover's and amplitude amplification algorithms.

\section{Implementing the Algorithm: Circuit-Level Design}\label{sec6}
\begin{figure}[H]
        \centering
        \fbox{\includegraphics[width=2.5cm]{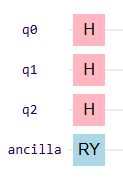}}
        \caption{Circuit for generating the initial state of the system.}
        \label{fig:detreministic_init}
    \end{figure}

\begin{figure}[H]
        \centering
        \fbox{\includegraphics[width=0.96\textwidth]{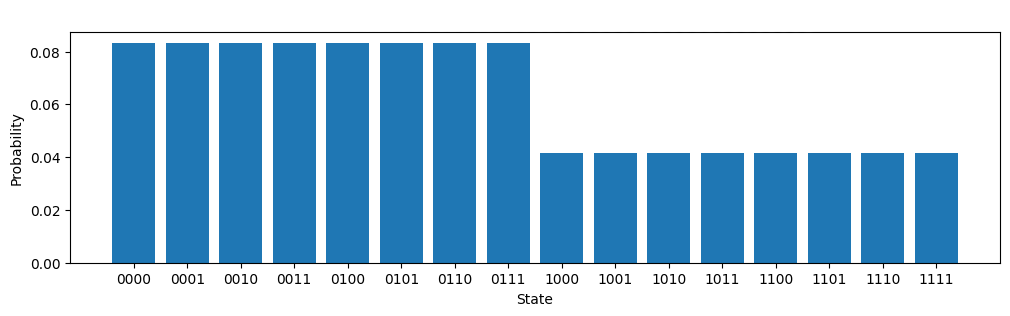}}
        \caption{Probability of measuring the states after initialization of the $ancilla$ qubit.}
        \label{fig:detreministic_init_graph}
    \end{figure}

\begin{figure}[H]
        \centering
        \fbox{\includegraphics[width=0.96\textwidth]{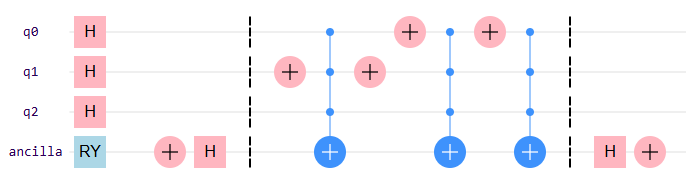}}
        \caption{Circuit for marking the new target states ($|\Psi_{new\_target}\rangle$).}
        \label{fig:deterministic_oracle}
    \end{figure}

\begin{figure}[H]
        \centering
        \fbox{\includegraphics[width=0.96\textwidth]{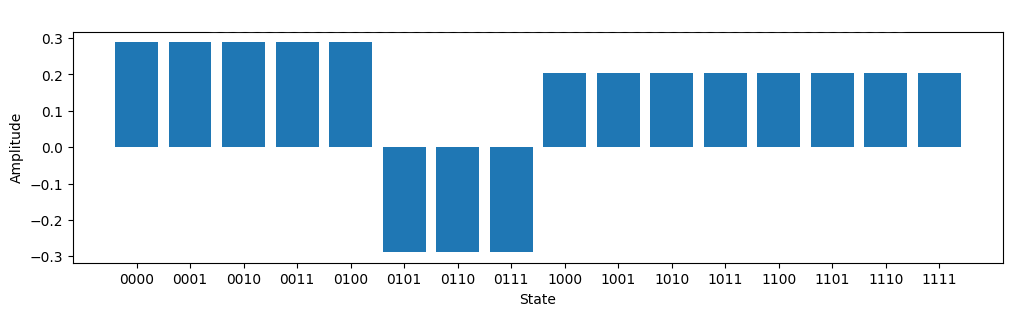}}
        \caption{Probability amplitude distribution of the states after the application of the oracle (negative amplitude of a particular state depicts the phase inversion of $\pi$).}
        \label{fig:deterministic_oracle_graph}
    \end{figure}

\begin{figure}[H]
        \centering
        \fbox{\includegraphics[width=0.96\textwidth]{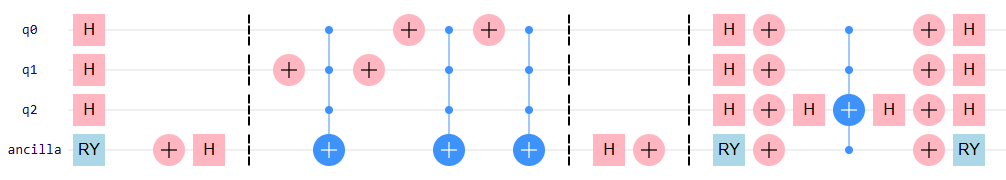}}
        \caption{Overall circuit for deterministic quantum search}
        \label{fig:deterministic_diffusion}
    \end{figure}

\begin{figure}[H]
        \centering
        \fbox{\includegraphics[width=0.96\textwidth]{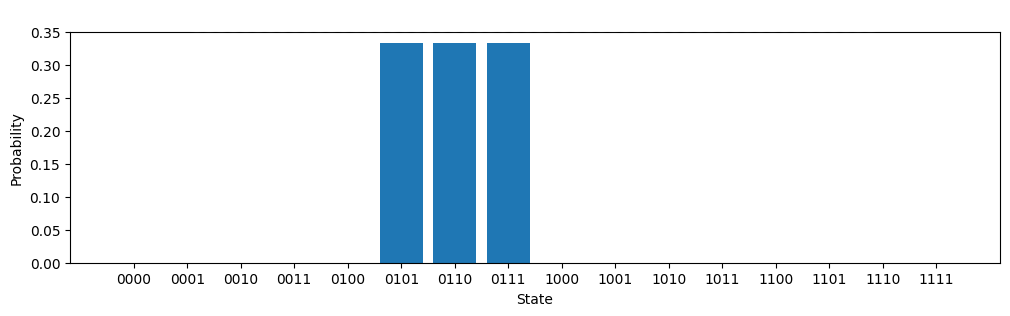}}
        \caption{Probability distribution of the states after applying diffusion operator algorithm.}
        \label{fig:deterministic_diffusion_graph}
    \end{figure}

\begin{figure}
        \centering
            \begin{subfigure}{1\textwidth}
                \includegraphics[width=0.94\textwidth]{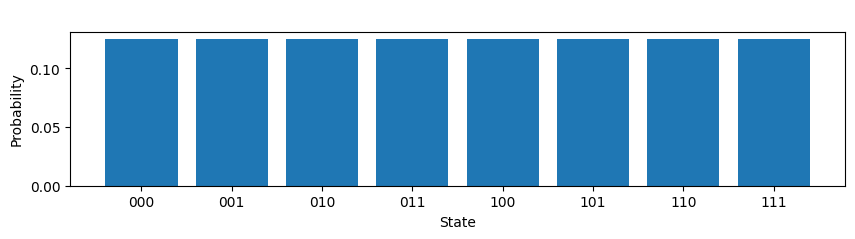}
                \caption{Probability of measuring the states in initial distribution of the system.}
                \label{fig:output}
            \end{subfigure}
        \vfill
            \begin{subfigure}{1\textwidth}
                \includegraphics[width=0.94\textwidth]{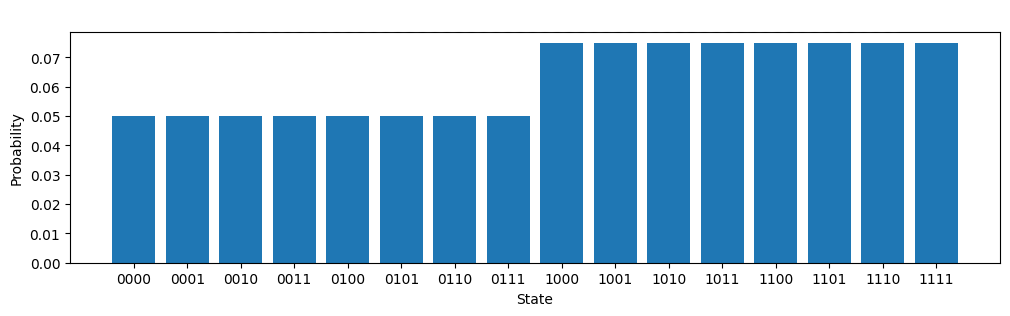}
                \caption{Probability of measuring the states after the initialization of the $ancilla$ qubit.}
                \label{fig:50_initial}
            \end{subfigure}
        \vfill
            \begin{subfigure}{1\textwidth}
                \includegraphics[width=0.94\textwidth]{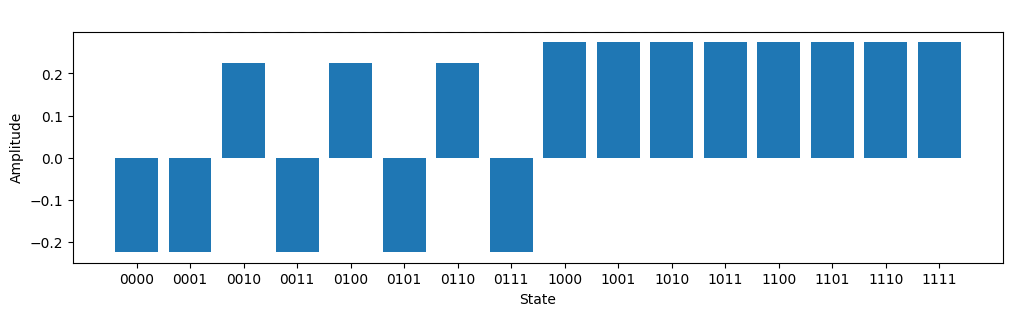}
                \caption{Amplitude distribution of the states after the application of oracle (negative amplitude depicts the phase inversion of $\pi$ of that particular state).}
                \label{fig:50_oracle}
            \end{subfigure}
        \vfill
            \begin{subfigure}{1\textwidth}
                \includegraphics[width=0.94\textwidth]{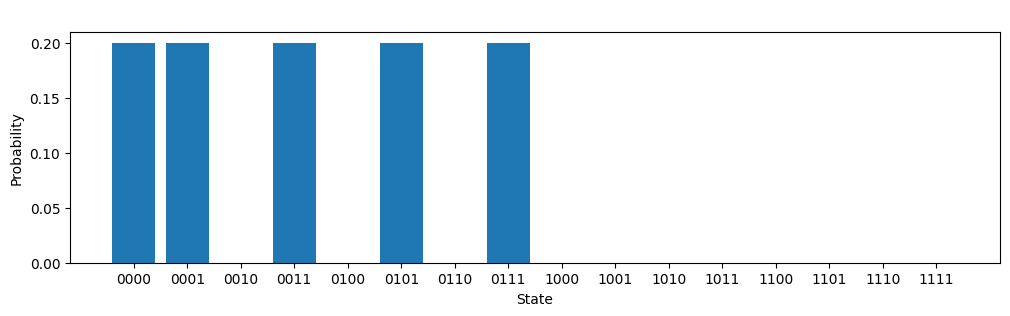}
                \caption{Probability of measuring the states after the deterministic quantum search algorithm is applied.}
                \label{fig:50_final}
            \end{subfigure}
        \caption{Probability distribution at different stages of the deterministic quantum search algorithm}
        \label{fig:subfigures4_new}
    \end{figure}

    This section provides a detailed analysis and circuit implementation for the proposed deterministic quantum search algorithm. To illustrate this, consider a search problem with a uniform superposition of 8 states:
    $$|\Psi\rangle = \frac{1}{\sqrt{8}}\sum_{i = 0}^{7}|i\rangle$$
    The objective is to search for the specific target states $|101\rangle$, $|110\rangle$ and $|111\rangle$ states. If the Grover's Algorithm is applied to search for these target states, the number of iterations calculated according to Eqn. \ref{eq:k} is approximately 0.6917, which is typically rounded to 1. This single iteration yields a success probability of approximately 0.84375, as calculated from Eqn. \ref{eq:p_success}.\\\\
    By using the deterministic quantum search algorithm discussed in this article the $ancilla$ qubit in the circuit is initialized with $R_y(\phi)$ rotational gate. The angle of rotation ($\phi$) is calculated from Eqn. \ref{eq: theta} to be approximately 1.2309 radians. The circuit is initialized as shown in Fig. \ref{fig:detreministic_init}, and the resulting probability distribution is plotted in Fig. \ref{fig:detreministic_init_graph}.
    \noindent Following initialization, the oracle is applied to mark the target states. As depicted in Fig. \ref{fig:deterministic_oracle}, the oracle marks the target states $|101\rangle, |110\rangle, |111\rangle$ only when the $ancilla$ qubit is in state $|0\rangle$. This results in the oracle marking the combined states $(|0101\rangle, |0110\rangle, |0111\rangle)$ by adding a phase of $\pi$. Consequently, after the oracle application, the probability amplitudes of these target states acquire a negative phase, while their magnitudes remain unchanged, as represented in Fig. \ref{fig:deterministic_oracle_graph}.
    \noindent Following this, the diffusion operator is applied to increase the probability of the marked target states. The overall circuit representation of this deterministic search algorithm is given in Fig. \ref{fig:deterministic_diffusion}.
    \noindent After applying the diffusion operator, the target states are assured to be measured with complete certainty. The probability distribution of the state of the system after diffusion operator is shown in Fig. \ref{fig:deterministic_diffusion_graph}. This illustration demonstrates that by using the proposed deterministic approach with an $ancilla$ qubit, target states can be obtained deterministically from an arbitrary initial state.
    \noindent To illustrate the advantage of this deterministic quantum search algorithm with the initial success probability being greater than $0.5$, consider the same search space as in the previous example, where there is a uniform superposition of $8$ states, represented in Fig. \ref{fig:output}. Let the target states in this case be $|000\rangle, |001\rangle, |011\rangle, |101\rangle, |111\rangle$. Hence, the initial probability of measuring the target states is $5/8 = 0.625$, which is more than $0.5$. As explained in Section \ref{sec2}, the Grover's algorithm and amplitude amplification algorithm do not offer a quantum advantage in such cases. The deterministic quantum search algorithm in this case, makes use of the $ancilla$ qubit initialized with $R_y(\phi)$ rotation gate. The probability distribution after initialization is shown in Fig. \ref{fig:50_initial}.
    \noindent The oracle marks the target states $|0000\rangle, |0001\rangle, |0011\rangle, |0101\rangle, |0111\rangle$ - with the $ancilla$ qubit being in state $|0\rangle$. This state of the system is represented in Fig. \ref{fig:50_oracle}. Applying the diffusion operator amplifies the probability amplitudes of these target states, leading to deterministic results, as illustrated in Fig. \ref{fig:50_final}, demonstrating $100\%$ accuracy even when the initial success probability is greater than $0.5$.

    \subsection*{Query and Circuit Complexity}
 
        The complexity of the proposed algorithm is analysed along four dimensions.
        
        \paragraph{(a) Oracle query complexity.}
        Each application of the controlled oracle is implemented by surrounding the standard black-box oracle $S_x$ with single-qubit gates (Pauli-$X$ and Hadamard gates) acting on the ancilla. Importantly, this construction invokes the oracle $S_x$ exactly once per iteration. Therefore, in the standard query model, each iteration counts as a single oracle query.\\\\
        The total number of oracle calls is at most $\lfloor k \rfloor + 1$, i.e., at most one call beyond the optimal (non-integer) iteration count. Hence, the oracle query complexity is $O(\sqrt{{N}/{M}})$ which matches the asymptotic complexity of standard Grover's and amplitude amplification algorithms.

        \paragraph{(b) Gate complexity.}
        Each iteration applies:\\
        (i)~a constant-size ancilla wrapper (two X and two H gates).\\
        (ii)~the oracle $S_x$ (gate count depends on the specific
            oracle implementation, denoted $|S_x|$).\\
        (iii)~the diffusion operator $D_f$ on the $(n+1)$-qubit system : $2(n+1)$ Hadamard gates plus one multi-controlled phase gate, i.e.,\ $O(n^2)$ gates via standard decomposition (adding a overhead of $O(n))$.\\
        (iv)~a one-time ancilla initialization via an $R_y(\phi)$ gate.\\
        Thus, the total gate complexity scales as $O(k' \cdot (n^2 + |S_x|)) = O((n^2  + |S_x|)*\sqrt{N/M})$, matching standard Grover's algorithm up to a overhead of $O(n)$ gates per iteration.
        
        \paragraph{(c) Circuit depth.}
        Each iteration has $D_{S_x}$, where $D_{S_x}$ is the depth of the oracle $S_x$. The ancilla wrapper contributes only constant additional depth and can be partially parallelized.
        The diffusion operator $D_f$ adds $O((\log n)^2)$ depth using
        a standard decomposition.
        Thus, the overall circuit depth is
        $O\bigl(k' \cdot (D_{S_x} + (\log n)^2)\bigr)$, which remains in the same asymptotic order as standard Grover search.
        
        \paragraph{(d) Ancilla resource overhead.}
        The proposed method does not requires any additional qubit beyond the $n$ data qubits and $1$ ancilla in the standard Grover's or amplitude amplification algorithm.
        
\section{Searching Multiple Targets in Definite Number of Steps}\label{sec7}
\begin{figure}[h]
        \centering
            \fbox{\includegraphics[width=\textwidth]{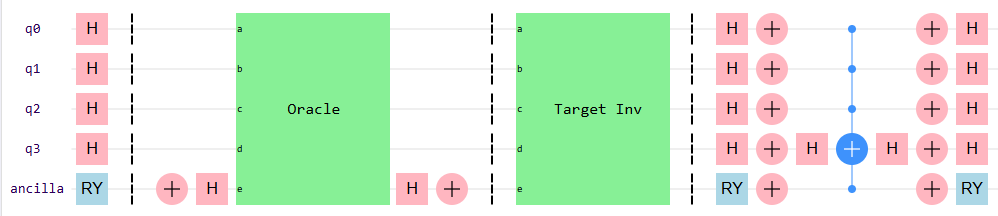}}
            \caption{Single iteration of the circuit for searching multiple target states in definite number of steps.(Target Inv is an operator to mark the previously measured target state(s) and has to be modified after measuring each target state).}
            \label{fig:multi_circ}
    \end{figure}

\begin{figure}
        \centering
        \begin{subfigure}[b]{0.25\textwidth}
            \fbox{\includegraphics[width=\textwidth]{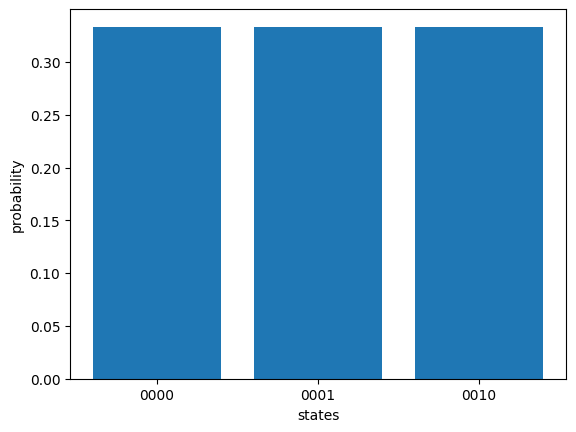}}
            \caption{Probability distribution after the $1^{st}$ execution of algorithm.}
        \end{subfigure}
        \hfill
        \begin{subfigure}[b]{0.25\textwidth}
            \fbox{\includegraphics[width=\textwidth]{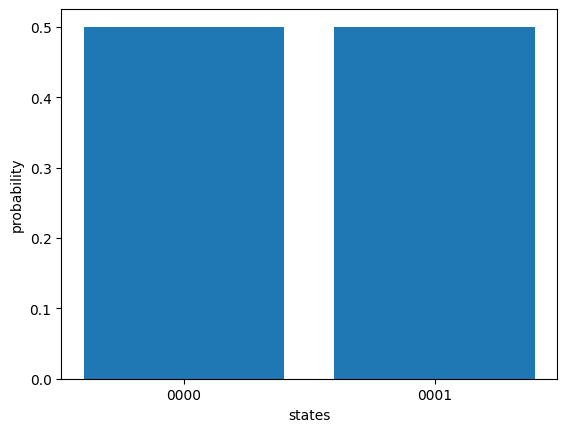}}
            \caption{Probability distribution after the $2^{nd}$ execution of algorithm.}
        \end{subfigure}
        \hfill
        \begin{subfigure}[b]{0.25
        \textwidth}
            \fbox{\includegraphics[width=\textwidth]{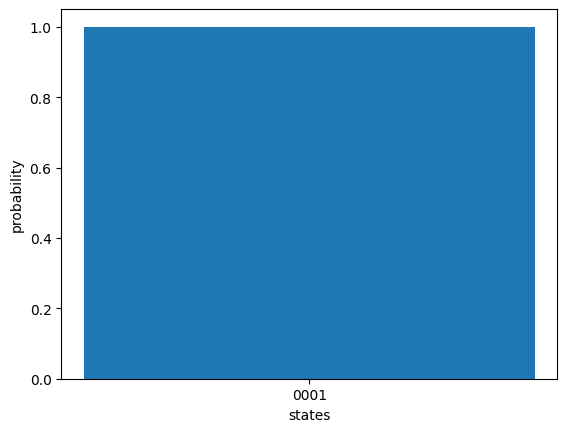}}
            \caption{Probability distribution after the $3^{rd}$ execution of algorithm.}
        \end{subfigure}
        \caption{Probability distribution after each execution of deterministic quantum search algorithm.}
        \label{fig:multiple_search}
    \end{figure}

    While searching for multiple target states ($M$), the Grover's and amplitude amplification algorithms produce a superposition of all target states. The measurement collapses the state of the system to any one of the target states. Hence, to obtain all target states, multiple instances of the algorithm are required to be executed. Even after multiple executions, the state of the system may end up with same target state. This might lead to ambiguity with respect to the number of times the algorithm is required to be executed to obtain all the target states. Further, there may be chances that the state of the system may collapse to some non-target state, due to the probabilistic nature of these algorithms.\\\\
    With the deterministic approach discussed in this article, the target state is obtained with certainty. Owing to the deterministic nature of this approach the target state is guaranteed to be obtained with a single execution of the algorithm. To obtain the other target state(s), the subsequent execution of the algorithm need not consider the target states which were already measured in previous instances. This is achieved by designing an oracle which only marks the remaining target state(s). However, this oracle needs to be modified for each execution of the algorithm.\\\\
    In order to overcome this challenge, a dynamic operator (Target Inv) can be designed and applied as depicted in Fig. \ref{fig:multi_circ}. This Target Inv operator, when applied after the oracle, marks the target states that are obtained in the previous executions, by applying a phase change of $\pi$. This results in an overall phase change of $2\pi$ ($\pi$ due to oracle and $\pi$ by Target Inv operator) of the earlier obtained target states, which is equivalent to a no phase change. The multi-target procedure employs an adaptive oracle construction in which, after each execution, previously measured target states are excluded using a $\text{Target Inv}$ operator. The measured target state is stored in classical memory and the corresponding $\text{Target Inv}$ circuit is updated incrementally. Each update requires $O(\log N)$ classical gate operations, corresponding to the number of qubits representing the search space. Over $M$ executions, the total classical overhead is $O(M \log N)$.\\\\
    By repeating this process $M$ times all the target states can be obtained with $100\%$ certainty in just $M$ executions of the algorithm.
    \begin{figure}
        \centering
            \fbox{\includegraphics[width=0.4\textwidth]{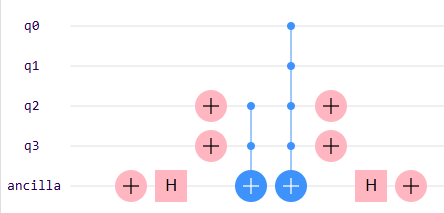}}
            \caption{Controlled oracle for marking $|0000\rangle, |0001\rangle, |0010\rangle$ states}
            \label{fig:multi_oracle}
    \end{figure}
    \noindent The working of the method discussed in this section can be understood by considering an example where there is uniform superposition of $16$ states, i.e. $|0000\rangle, |0001\rangle, |0010\rangle, ..., |1111\rangle$ and let the target states be $|0000\rangle, |0001\rangle, |0010\rangle$. The oracle circuit to mark these states is given in Fig.\ref{fig:multi_oracle}. The algorithm is to be executed $3$ times if the deterministic quantum search discussed in this article is used. The Target Inv operator, applied after each oracle call, would effectively ignore the target states found in earlier executions. The probability distribution of states after each execution is shown in Fig. \ref{fig:multiple_search}.\\
    \\
    \noindent In the first execution of the algorithm, the state $|0010\rangle$ was measured. Hence in the next execution $|0010\rangle$ state was not considered as a target state. In the $2^{nd}$ execution, $|0000\rangle$ was measured and subsequently in the $3^{rd}$ execution $|0001\rangle$ was measured. The query complexity for searching $M$ targets in a search space of size $N$ is O($M\sqrt{N}$), as each of the $M$ deterministic searches takes O($\sqrt{N}$) query. In this way using this approach, all the target states can be measured deterministically with a definite number of executions.\\\\


    

\section{Conclusion}\label{sec9}
    The proposed deterministic quantum search algorithm demonstrates substantial advancement within the realm of quantum computing, offering deterministic results compared to the existing probabilistic quantum search algorithms. Despite its deterministic nature, this approach maintains the same order of circuit depth as Grover's and amplitude amplification algorithms. The majority of the circuit construction utilizes existing components from these algorithms; the initialization operator is built by applying the $R_y(\phi)$ operator on the $ancilla$ qubit and the oracle and diffusion operators are invoked in a controlled manner using the ancilla qubit.\\ \\
    In Grover's and amplitude amplification algorithms, when the number of iterations yields a non-integer value, it is rounded to the nearest natural number by considering either the lower or upper bound. This rounding leads to an undershoot or an overshoot, resulting in probabilistic outcomes. In contrast, the deterministic quantum search approach always considers the upper bound of the calculated number of iterations and requires at most one additional iteration to guarantee certainty.\\\\
    These modifications ensure that the inherent limitations of Grover's and amplitude amplification algorithms are overcome, providing deterministic results and offering a quantum advantage even when the initial success probability is greater than $0.5$.\\\\
    The proposed deterministic approach requires knowledge of the initial success probability $p$. Extending the framework to scenarios where $p$ is unknown presents a direction for future work. In such settings, integrating the method with quantum counting or amplitude estimation techniques to infer $p$ could enable a fully adaptive deterministic search procedure.\\\\
    The explicit circuit design of the proposed deterministic quantum search algorithm, as presented in this work, demonstrates its implementation across the entire range of initial success probabilities. Moreover, a method for searching multiple targets within a fixed number of steps is introduced. This work also provides an analysis of the algorithm’s query complexity.\\\\
    In conclusion, the approach discussed in this article represents an advancement in the field of quantum computing by mitigating the probabilistic nature of existing quantum search algorithms and guaranteeing the successful identification of target states with certainty. The proposed algorithm not only enhances the reliability of quantum search algorithms, but also opens new avenues for applying quantum computing in areas requiring precise and deterministic results.


\section*{Acknowledgment}
    The quantum circuit implementations and corresponding results presented in this work were obtained using \textbf{Qniverse : A Unified Quantum Computing Platform} \cite{Qniverse}. This research was supported by the Ministry of Electronics and Information Technology (MeitY), Government of India, under the project titled \emph{“HPC-based Quantum Accelerators for Enabling Quantum Computing on Supercomputers”}, Grant No. 4(3)/2022-ITEA.


\begin{appendices}

    \section{Working of Oracle in Grover's Algorithm}\label{secA1}
        This appendix details the mathematical mechanics of the oracle operator within the standard Grover's algorithm, illustrated by the generic quantum circuit schematic illustrated in Fig. \ref{fig:internal oracle}. Specifically, it demonstrates how the standard oracle applies a phase flip exclusively to the target states. These foundational principles serve as the baseline upon which the components of the proposed deterministic quantum search is built. To demonstrate this, a step-by-step evolution of the quantum state as it passes through the circuit is traced.\\
        \begin{figure}[h]
        \centering
            \definecolor{myblue}{RGB}{173, 216, 230}
            \definecolor{myred}{RGB}{255, 182, 193}
            \definecolor{mygreen}{RGB}{144, 238, 144}
            \definecolor{mygrey}{RGB}{151, 151, 151}
            \fbox{\begin{quantikz}
                \lstick{$\ket{0}^{\otimes n}$} 
                    & \qwbundle{n}\slice[style=black]{\ket{\Psi_0}}
                    & \gate[style={fill = myred}]{H^{\otimes n}}\slice[style=black]{\ket{\Psi_1}}
                    & \gate[style={fill=mygreen},label style=black, 2][3cm]{S_x}\gateinput{$x$}\gateoutput{$x$}\slice[style=black]{\ket{\Psi_2}}
                    & \\
                \lstick{ancilla $\ket{-}$} 
                    &                       
                    &
                    & \gateinput{$y$}\gateoutput{$y\oplus f(x)$}
                    &
            \end{quantikz}}
        \caption{Generalized quantum circuit illustrating the oracle operation in Grover's algorithm. The $ancilla$ qubit is initialized in the $|-\rangle$ state, while the first register of $n$ qubits is initialize using $H^{\otimes n}$. Sx acts on inputs $x$ and $y$ and produces outputs $x$ and $y\oplus f(x)$, where $f(x)$ is the function defined in the problem statement.}
        \label{fig:internal oracle}
    \end{figure}\\
    The initial state $\ket{\Psi}$ of the system is given by
    $$\ket{\Psi_0} = \ket{0}^{\otimes n} \otimes \ket{-}$$
    where:
    $$|-\rangle = \frac{|0\rangle - |1\rangle}{\sqrt{2}}$$
    After initializing the first quantum register of $n$ qubits using the Hadamard operator ($H^{\otimes n}$), the state of the system becomes
    $$\ket{\Psi_1} = \frac{1}{\sqrt{2^n}}\sum_{x\rightarrow \{0,1\}^n} \ket{x} \otimes \ket{-}$$
    The oracle operates on this uniform superposition of all possible states of $n$ qubits register and the state of ancilla qubit ($\ket{-}$ state). The state of the system changes as follows:
    $$\ket{\Psi_2} = \frac{1}{\sqrt{2^n}}\sum_{x\rightarrow \{0,1\}^n} \ket{x} \otimes \frac{1}{\sqrt{2}}\Big(\ket{0 \oplus f(x)} - \ket{1 \oplus f(x)}\Big)$$
    Thus, in general, the state can be written as
    $$\ket{\Psi_2} = \frac{1}{\sqrt{2^n}}\sum_{x\rightarrow \{0,1\}^n} (-1)^{f(x)}\ket{x} \otimes \ket
    {-}$$
    Therefore, the oracle operator applies a phase flip only to the target states (states $x$ for which $f(x) = 1$).

    \section{Proof of Existence of Modified Success Probability ($p'$)}\label{secA3}

    \begin{theorem}[Existence of Modified Success Probability for Deterministic Amplification]
Let $0 < p < 1$ be the initial success probability in an amplitude amplification procedure, and let
\[
\theta = sin^{-1}(\sqrt{p}) \in (0, \tfrac{\pi}{2}).
\]
Define the ideal (real-valued) Grover iteration count as
\[
k = \frac{\pi}{4\theta} - \frac{1}{2}.
\]
If $k\notin \mathbb{Z}$, then there exists a modified success probability $p' < p$, corresponding to an angle $\theta' < \theta$, and an integer $k' = \lfloor k \rfloor + 1$, such that after $k'$ iterations of amplitude amplification, the success probability becomes exactly $1$.
\end{theorem}

\begin{proof}
In amplitude amplification, the state evolution occurs in a two-dimensional subspace spanned by the target and non-target states. After $k$ iterations, the success probability is given by
\[
p_{\text{success}} = sin^2((2k+1)\theta).
\]
The condition for deterministic success is
\[
(2k'+1)\theta' = \frac{\pi}{2}
\]
for some integer $k'$ and angle $\theta'$.\\\\
For a given initial $\theta$, the ideal number of iterations required to reach $\pi/2$ is
\[
k = \frac{\pi}{4\theta} - \frac{1}{2}.
\]
If $k \notin \mathbb{Z}$, then choosing $k' = \lfloor k \rfloor + 1$ yields
\[
(2k'+1)\theta > \frac{\pi}{2},
\]
leading to overshoot.\\\\
Since $k' > k$, we have
\[
2k'+1 > 2k+1 = \frac{\pi}{2\theta},
\]
which implies\\
\[
\frac{\pi}{2(2k'+1)} < \theta.
\]\\
Defining\\
\[
\theta' = \frac{\pi}{2(2k'+1)}.
\]\\
So that $(2k+1)\theta'= \pi/2$ holds by construction.\\\\
Thus, there exist $\theta' < \theta$, 
Finally, substituting $\theta'$ into the success probability expression:
\[
p_{\text{success}} = sin^2((2k'+1)\theta') = sin^2\left(\frac{\pi}{2}\right) = 1.
\]
and defining
\[
p' = sin^2(\theta'),
\]\\
and noting that $sin^2(\theta')$ is strictly increasing on $[0,\pi/2]$, it follows that $p'<p$.
Since $p' < p$, we have 
\[
|\alpha|^2 = \frac{p'}{p} \in (0,1),
\]
and hence there exists a valid ancilla rotation $R_y(\phi)$ such that
\[
|\alpha|^2 = cos^2\!\left(\frac{\phi}{2}\right).
\]
Therefore, the modified success probability $p'$ is physically realizable.\\ \\
Thus, for any $0 < p < 1$ with non-integer $k$, there exists a modified success probability $p' < p$ such that deterministic success is achieved after $k'$ iterations.
\renewcommand{\qedsymbol}{}
\end{proof}
\noindent This result formalizes the key idea that deterministic quantum search can always be achieved by appropriately rescaling the initial success probability, rather than modifying the phase structure of the Grover operator.

    \section{Controlled Phase Marking Using a Black-Box Oracle} \label{secA2}

    \textbf{Lemma 1 (Controlled Phase Kickback via Ancilla Rotation)}\\
    Let $S_x$ be a standard black-box oracle defined as:
    $$S_x |x\rangle |y\rangle = |x\rangle |y \oplus f(x)\rangle$$
    where $f(x) \in \{0,1\}$. By surrounding $S_x$ with appropriate single-qubit operations acting only on an $ancilla$ qubit, the oracle can be transformed into a controlled phase-marking operation such that a phase factor $(-1)^{f(x)}$ is conditionally applied based on the $ancilla$ state, while preserving the internal black-box nature of $S_x$.\\ \\
    \textbf{Proof}\\ \\
    To demonstrate this, consider the circuit illustrated in Fig. \ref{fig:controlled_oracle}. We analyze the state evolution of the system step by step.\\
    \begin{figure}[h]
        \centering
            \definecolor{myblue}{RGB}{173, 216, 230}
            \definecolor{myred}{RGB}{255, 182, 193}
            \definecolor{mygreen}{RGB}{144, 238, 144}
            \definecolor{mygrey}{RGB}{151, 151, 151}
            \fbox{\begin{quantikz}
                \lstick{$\ket{0}^{\otimes n}$} 
                    & \qwbundle{n}          
                    & \gate[style={fill=mygreen},label style=black]{A}\slice[style=black, label style={yshift=0.3cm}]{\ket{\Psi_0}}             
                    &\slice[style=black, label style={yshift=0.3cm}]{\ket{\Psi_1}}
                    \gategroup[wires=2, steps=5, style={dashed, rounded corners, inner sep=9pt}, label style={label position=below, anchor=north, yshift=-0.3cm}]{Controlled Oracle} 
                    & \slice[style=black, label style={yshift=0.3cm}]{\ket{\Psi_2}}
                    & \gate[style={fill=mygreen},label style=black, 2][3cm]{S_x}\gateinput{$x$}\gateoutput{$x$}\slice[style=black, label style={yshift=0.3cm}]{\ket{\Psi_3}}         
                    & \slice[style=black, label style={yshift=0.3cm}]{\ket{\Psi_4}}
                    & \slice[style=black, label style={yshift=0.3cm}]{\ket{\Psi_5}}
                    & \\           
                \lstick{ancilla $\ket{0}$} 
                    &                       
                    & \gate[style={fill=myblue},label style=black]{R_y(\phi)}      
                    & \gate[style={fill=myred, circle, inner sep=-2pt},label style=black]{+}
                    & \gate[style={fill=myred},label style=black]{H}
                    & \gateinput{$y$}\gateoutput{$y\oplus f(x)$}          
                    & \gate[style={fill=myred},label style=black]{H}
                    & \gate[style={fill=myred, circle, inner sep=-2pt},label style=black]{+}
                    &
            \end{quantikz}}
        \caption{By placing $X$ and $H$ gates on the $ancilla$ qubit around the black-box oracle ($S_x$), the circuit triggers a phase inversion on the first quantum register of $n$ qubits exclusively when the $ancilla$ qubit is in state $|0\rangle$ and $f(x)=1$. This implements the controlled operation while strictly preserving the unobservable, black-box nature of $S_x$}
        \label{fig:controlled_oracle}
    \end{figure}\\
    Let the $n$-qubit quantum register be in a computational basis state $|x\rangle$ and let the $ancilla$ qubit be initialized in the superposition:
    $$\alpha|0\rangle + \beta|1\rangle$$\\
    generated by applying an $R_y(\phi)$ rotation to the state $|0\rangle$.\\
    The initial state entering the controlled oracle block is:
    $$\Psi_0 = |x\rangle \otimes (\alpha|0\rangle + \beta|1\rangle)$$\\
    Although the derivation is presented for a computational basis state $|x\rangle$, the result extends directly to arbitrary superpositions of basis states due to the linearity of quantum operations.\\ \\
    \textbf{Step 1: Application of the Pauli-X Gate}\\
    The Pauli-X gate flips the $ancilla$ qubit, producing:
    $$\Psi_1 = (I^{\otimes n} \otimes X)\Psi_0 = |x\rangle \otimes (\alpha|1\rangle + \beta|0\rangle)$$\\
    \textbf{Step 2: Application of the Hadamard Gate}\\
    Applying the Hadamard gate transforms the $ancilla$ qubit into the superposition basis:
    $$\Psi_2 = (I^{\otimes n} \otimes H)\Psi_1 = |x\rangle \otimes (\alpha|-\rangle + \beta|+\rangle)$$
    where:
    $$|+\rangle = \frac{|0\rangle + |1\rangle}{\sqrt{2}}$$\\
    \textbf{Step 3: Application of the Black-Box Oracle}\\
    The oracle evaluates the Boolean function $f(x)$ and flips the $ancilla$ qubit accordingly:
    $$\Psi_3 = S_x\Psi_2 = |x\rangle \otimes \left[ \alpha(-1)^{f(x)} |-\rangle + \beta|+\rangle \right]$$\\
    \textbf{Step 4: Second Hadamard Gate}\\
    Applying the Hadamard gate again transforms the $ancilla$ qubit back into the computational basis:
    $$\Psi_4 = (I^{\otimes n} \otimes H)\Psi_3 = |x\rangle \otimes \left[ \alpha(-1)^{f(x)} |1\rangle + \beta|0\rangle \right]$$\\
    \textbf{Step 5: Second Pauli-X Gate}\\
    Finally, the Pauli-X gate restores the logical orientation of the $ancilla$ qubit:
    $$\Psi_5 = (I^{\otimes n} \otimes X)\Psi_4 = |x\rangle \otimes \left[ \alpha(-1)^{f(x)} |0\rangle + \beta|1\rangle \right]$$\\
    \textbf{Conclusion}\\
    Factoring the final state $\Psi_5$ shows that the phase factor $(-1)^{f(x)}$ is applied to the $n$-qubit register state $|x\rangle$ if and only if the $ancilla$ qubit evaluates to the state $|0\rangle$.\\ \\
    Therefore, the required controlled phase-marking operation can be implemented by surrounding the black-box oracle $S_x$ with standard single-qubit gates acting only on the $ancilla$ qubit without revealing or modifying the internal structure of the oracle.

    \vspace{1em}
    \noindent\textbf{Operator Representation of the Controlled Oracle}\\
    The sequence of operations applied to the $ancilla$ qubit can be written compactly as:
    $$U = (I^{\otimes n} \otimes X)(I^{\otimes n} \otimes H) S_x (I^{\otimes n} \otimes H)(I^{\otimes n} \otimes X)$$\\
    Applying this operator to the state $$|x\rangle \otimes (\alpha|0\rangle + \beta|1\rangle)$$ yields:
    $$U(|x\rangle \otimes (\alpha|0\rangle + \beta|1\rangle)) = |x\rangle \otimes \left[ \alpha(-1)^{f(x)}|0\rangle + \beta|1\rangle \right]$$\\
    Thus the composite circuit effectively realizes a controlled phase oracle:
    $$|x\rangle \rightarrow (-1)^{f(x)} |x\rangle$$
    conditioned on the $ancilla$ qubit being in state $|0\rangle$. This confirms that the phase marking can be achieved without decomposing or modifying the internal structure of the black-box oracle $S_x$.
\end{appendices}

\section*{Availability of Data and Materials}
    The complete source code and Jupyter notebooks used to generate the quantum circuit implementations and probability distributions analysed in this study are publicly available in the Deterministic Quantum Search GitHub repository: https://github.com/C-DAC-Bengaluru/Qniverse/tree/main/quantum-algorithm-developments/Deterministic\%20Quantum\%20Search. The algorithmic implementation was developed and evaluated utilizing Qniverse: A Unified Quantum Computing Platform \cite{Qniverse}.

\printbibliography

\end{document}